%% file: main.tex
\documentclass[11pt]{article}
\usepackage[margin=1.0in]{geometry}
\usepackage[english]{babel}

\usepackage{ifpdf}
\ifpdf    
\usepackage[pdftex,colorlinks,linkcolor=blue,citecolor=blue,filecolor=blue,urlcolor=blue]{hyperref}
\else    
\usepackage[hypertex,colorlinks,linkcolor=blue,citecolor=blue,filecolor=blue,urlcolor=blue]{hyperref}
\fi

\usepackage{chngcntr}
\usepackage{amsmath}
\usepackage{amssymb}
\usepackage{amsfonts}
\usepackage{amsthm}
\usepackage{dsfont}
\usepackage{algorithm}
\usepackage{algpseudocode}
\usepackage{enumerate}
\usepackage{color}
\usepackage{multirow}
\usepackage{xspace}
\usepackage{mathtools}
\usepackage{thmtools}
\usepackage{thm-restate}
\usepackage{tikz}
\usepackage{etoolbox}

\usetikzlibrary{positioning}
\usetikzlibrary{decorations.text}
\usetikzlibrary{decorations.pathmorphing}
\usetikzlibrary[topaths]

\newtheorem{theorem}{Theorem}[section]
\newtheorem*{theorem*}{Theorem}
\newtheorem{lemma}[theorem]{Lemma}

\newtheorem{definition}[theorem]{Definition}

\newtheorem*{claim*}{Claim}

\newtheorem{proposition}[theorem]{Proposition}

\title{Matrix Norms in Data Streams: Faster, Multi-Pass and Row-Order}

\author{Vladimir Braverman\thanks{
Email: \texttt{vova@cs.jhu.edu}.
This material is based upon work supported by the NSF Grants IIS-1447639,  EAGER CCF-1650041, and CAREER CCF-1652257
}
\\ Johns Hopkins University
\and Stephen R.\ Chestnut\thanks{
Email: \texttt{stephenc@ethz.ch.}
}
\\ ETH Zurich 
\and
Robert Krauthgamer\thanks{%
  Email: \texttt{robert.krauthgamer@weizmann.ac.il}. 
  Work supported in part by the Israel Science Foundation grant \#897/13.
}
\\ Weizmann Institute of Science
\and
Yi Li\thanks{%
  Email: \texttt{leeyi@umich.edu}. 
}\\
 Nanyang Technological University
\and 
David P. Woodruff\thanks{
Email: \texttt{dwoodruf@cs.cmu.edu}.
}
\\ Carnegie Mellon University
\and 
Lin F. Yang\thanks{
Email: \texttt{lin.yang@princeton.edu}.
This material is based upon work supported by the NSF Grant IIS-1447639.
Work was done while the author was in Johns Hopkins University.
}
\\ Princeton University
}

\newcommand{\calA}{\mathcal{A}}
\newcommand{\calB}{\mathcal{B}}

\newcommand{\calD}{\mathcal{D}}

\newcommand{\bbR}{\mathbb{R}}

\DeclareMathOperator{\diag}{diag}
\DeclareMathOperator{\poly}{poly}
\DeclareMathOperator{\polylog}{polylog}

\DeclareMathOperator{\tr}{Tr}

\newcommand{\E}{\mathbb{E}}

\newcommand{\R}{\mathbb{R}}

\newcommand{\zdsketch}{$ZD$-sketch}

\DeclareMathOperator{\EX}{{\mathbb E}}
\DeclareMathOperator{\ext}{\EX}
\DeclareMathOperator{\Var}{Var}

\newcommand\norm[1]{\Vert {#1}\Vert}

\newcommand\tuple[1]{\langle {#1}\rangle}
\providecommand{\eqdef}{\coloneqq}
\providecommand{\set}[1]{{\{#1\}}}
\newcommand{\indic}{\mathds{1}}
\newcommand\indice[1]{\indic_{\set{#1}}}
\newcommand{\eps}{\epsilon}
\newcommand{\tO}{\tilde{O}}
\providecommand{\ceil}[1]{\lceil #1 \rceil}

\newcommand{\Z}{\mathbb{Z}}

\def\compactify{\itemsep=0pt \topsep=0pt \partopsep=0pt \parsep=0pt}
\newcommand{\Holder}{H\"older\xspace}

\newcommand{\mathp}{$p$}

\begin{document}

\maketitle

\thispagestyle{empty}
\setcounter{page}{0}

\begin{abstract}
\input{abstract}

\end{abstract}

\clearpage
\input{intro}

\input{prelim}
\input{sec_schatten}
\input{row_order}
\input{row_order_ub}

{\small
\bibliographystyle{alphaurlinit}
\bibliography{ref}
}

\appendix
\counterwithin{theorem}{section}
\input{appendix}
\end{document}

%% file: abstract.tex
A central problem in data streams is to characterize which functions of an underlying frequency vector can be approximated efficiently. 
Recently there has been considerable effort in extending this problem to that of estimating functions of a matrix that is presented as a data-stream. 
This setting generalizes classical problems to the analogous ones for matrices. 
For example, instead of estimating frequent-item counts, 
we now wish to estimate ``frequent-direction'' counts. 
A related example is to estimate norms, which now correspond to estimating a vector norm on the singular values of the matrix. 
Despite recent efforts, the current understanding for such matrix problems 
is considerably weaker than that for vector problems. 

We study a number of aspects of estimating matrix norms in a stream that have not previously been considered: 
(1) multi-pass algorithms, (2) algorithms that see the underlying matrix one row at a time, and (3) time-efficient algorithms. 
Our multi-pass and row-order algorithms use less memory 
than what is provably required in the single-pass and entrywise-update models, 
and thus give separations between these models (in terms of memory).
Moreover, all of our algorithms are considerably faster than previous ones. 
We also prove a number of lower bounds, and obtain for instance, 
a near-complete characterization of the memory required
of row-order algorithms for estimating Schatten $p$-norms of sparse matrices.

%% file: intro.tex
\section{Introduction}

Modern datasets, from text documents and images to social graphs,
are often represented as a large matrix $A\in\R^{m\times n}$.
In many application domains, including database queries, 
data mining, network transactions and sensor networks
(see e.g.\ \cite{liberty2013simple,WLLSDW16,HKS16} for recent examples), 
the input matrix $A$ is presented to the algorithm as a data stream,
i.e., a sequence of items/updates that can take several forms.
In the \emph{entry-wise (or insertion-only) model}, each item specifies $(i,j,A_{ij})$
and provides the value of one entry, in arbitrary order
(and the unspecified entries are set to $0$).
The \emph{row-order model} is similar,
except that the items follow the natural order
(sorted with $i$ as the primary key, and $j$ as the secondary one).
In the \emph{turnstile model}, each stream item has the form $(i,j,\delta)$
and represents an update $A_{ij}\gets A_{ij} + \delta$ for $\delta\in\R$
(after initializing $A$ to the all-zeros matrix).
These models capture different access patterns,
but all three can represent sparse matrices quite efficiently,
because zero entries are implicit.
As usual, the key parameters of an algorithm in the data-stream model
are its memory (also referred to as storage/space requirements) and its runtime
(per update and to report its output).

Many properties of a matrix are directly related to its spectral
characteristics, i.e., its singular values.
For example, the number of non-zero singular values is just the matrix rank, 
which determines the degrees of freedom of a corresponding linear system;
the maximum and minimum singular values of a matrix determine its condition
number, which in turn determines the hardness of many problems, such as
optimization problems; the leading singular values of a matrix determine how
well a matrix can be represented by the principal components; and so forth. 
It is generally hard to compute directly the singular values of a matrix,
especially in the streaming model, 
but luckily, the Schatten norms of the matrix can often be used 
as surrogates for its spectrum, see e.g.\ \cite{zhang2015distributed,
kong2016spectrum, di2016efficient, khetan2017spectrum}.
Formally, the \emph{Schatten $p$-norm} of a matrix $A\in\R^{m\times n}$ 
is defined, for every $p\ge 1$, as
 \[
 \|A\|_{S_p}\eqdef \Big(\sum_{j\geq1}\sigma_j^p\Big)^{1/p},
 \]
where $\sigma_1\geq \cdots\geq \sigma_{\min(m,n)}$ are the singular values
of $A$.
This definition naturally extends to all $0<p<1$ although then it is not a norm,
and also to $p=0,\infty$ by taking the limit. 
This is a very important family of matrix norms,
and includes as special cases 
the well-known trace/nuclear norm 
$\norm{A}_* = \sum_{j\geq 1}\sigma_j = \|A\|_{S_1}$, 
the Frobenius norm 
$\norm{A}_F 
 = \big(\sum_{j\geq1}\sigma_j^2\big)^{1/2}
 = \|A\|_{S_2}$, 
and the spectral/operator norm
$\|A\|_{op} = \sigma_1(A) = \norm{A}_{S_\infty}$.

We study algorithms that approximate the Schatten $p$-norm of a matrix $A$
presented in a data stream.
While this problem has attracted significant attention lately 
\cite{andoni2013eigenvalues,li2014sketching,
li2016approximating,li2016tight,li2017embeddings},
our results address three new aspects.
First, we design faster and more space-efficient \emph{multi-pass} algorithms.
Second, we consider the \emph{row-order model}, which is 
a common access pattern for matrix data (see, e.g.~\cite{liberty2013simple}).
Third, we design algorithms with faster \emph{update time and/or query time}.
The above three aspects were not considered previously for matrix norms, 
and our work opens the door for further diversification of prevailing models (and thereby of current algorithms).
In particular, our results can be applicable to classical scenarios, 
e.g., where data is stored on disk 
(or any media where a linear scan is much faster than random access),
and potentially lead to performance improvements in other such domains. 
In the next few subsections, we present our contributions in more detail.

\subsection{New Estimator for PSD Matrices (or Even $p$)}
\label{sec:IntroSp}
Our first results rely on a new method for estimating the Schatten $p$-norm
$\norm{A}_{S_p}$ of a positive semidefinite matrix (PSD) matrix $A\in\R^{n\times n}$ for integer $p\ge 2$.
This method yields two new streaming algorithms in the turnstile model,
which require, respectively, one pass and $\ceil{p/2}$ passes over the
input.
Both algorithms are at least as good as the previous ones in all three
standard performance measures of storage, update time, and query time;
and each algorithm offers significant improvements in two out of these
three.
Our one-pass algorithm achieves update time $O(1)$
compared with the previous $\poly(n)$,
and query time $O(n^{\omega(1-p/2)})$, where $\omega\leq 2.373$ is the
matrix multiplication exponent~\cite{le2014powers},
compared with the previous $n^{p-2}$.
And our multi-pass algorithm requires storage that is sublinear in $n$,
compared with $O(n)$ previously.
We note that if $p$ is even, then the above results extend to arbitrary
$A\in\R^{m\times n}$ (and not only PSD) by a standard argument.
A detailed comparison of the bounds is given in Table~\ref{tab:results},
and the results themselves appear in Section~\ref{sec:Sp}.

Throughout the paper, a matrix is called \emph{sparse} if it has at most 
$O(1)$ non-zero entries per row and per column. 
We write $\tilde O(f)$ as a shorthand for $O(f\cdot\log^{O(1)} f)$, 
and write $O_{a}(f)$ to indicate that the constant in $O$-notation depends on 
some parameter $a$.

\begin{table*}[t]
\begin{center}
\begin{tabular}{|lcllll|}
  \hline
  \multicolumn{6}{|l|}{Problem: Schatten $p$-norm of PSD $A$, integer $p\ge
2$ (or general $A$, even $p$)} \\
  \hline
  & passes & space & update time & query time & \\
  & $1$ & $\eps^{-2} n^{2-4/p}$ & $\eps^{-2} n^{2-4/p}$ &
$\epsilon^{-2}n^{p-2}$
  & \cite{li2014sketching}
  \\
  & $1$ & $\eps^{-2} n^{2-4/p}$ & $\eps^{-2}$ & $\eps^{-2}
n^{(1-2/p)\omega}$
  & Theorems~\ref{thm:Sp} and~\ref{thm:fast update algorithm general}
  \\
  & $\ceil{p/2}$ & $\epsilon^{-2}n$ & $\eps^{-2}$ & $\eps^{-2}n$
  & \cite[Theorem 6.1]{woodruff2014sketching}
  \\
  & $\ceil{p/2}$ & $\eps^{-2} n^{1-1/(p-1)}$ & $\eps^{-2}$ & $\eps^{-2}
n^{(1-1/(p-1))}$
  & Theorems~\ref{thm:multipass} and~\ref{thm:fast update algorithm general}
  \\
  \hline
\end{tabular}
\end{center}
\caption{Streaming algorithms for $(1+\eps)$-approximation of the Schatten $p$-norm of $A\in\R^{n\times n}$. 
The bounds for storage/time omit $O_p(1)$ factors, 
and count space in words.
}
\label{tab:results}
\end{table*}

\paragraph*{Techniques}
Our technical innovation is an unbiased estimator of $\tr(A^p)$
for a \emph{symmetric} (and not only PSD) matrix $A\in\R^{n\times n}$.
To see why this is useful,
denote the eigenvalues of $A$ by $\lambda_1\ge\cdots\ge \lambda_n$,
and observe that if $A$ is PSD (or alternatively if $p$ is even),
then
$\tr(A^p)=\sum_i \lambda_i^p = \sum_i \sigma_i(A)^p = \norm{A}_{S_p}^p$.
%
Our estimator has the form
\begin{equation} \label{eq:X}
  X\eqdef   \tr(G_1 AG_2^T G_2 AG_3^T\cdots G_pAG_1^T) ,
\end{equation}
where $G_i\in\R^{t\times n}$ are certain random matrices.
This estimator $X$ can be computed from the $p$ bilinear sketches
$\set{G_iAG_{i+1}^T}_{i\in[p]}$ by straightforward matrix multiplication, where $G_{p+1}\eqdef G_1$ by convention.
And if, say, $t=O(n^{1-2/p})$,
then each bilinear sketch has dimension $O(t^2)$ $= O(n^{2-4/p})$.
These determine the streaming algorithm's storage requirement and query
time,
and, if the matrices $\set{G_i}_{i\in[p]}$ have sparse columns,
the updates will be fast.

The main difficulty is to bound the estimator's variance,
which highly depends on the choice of the matrices $\set{G_i}_{i\in[p]}$.
The basics of this technique can be seen in the case $p=4$,
if the $G_i$'s satisfy the following definition.
\begin{definition} \label{def:JLT}
A random matrix $S\in\R^{t\times n}$ is called an \emph{$(\epsilon,\delta,d)$-Johnson-Lindenstrauss
Transformation (JLT)} if for every $V\subseteq\R^n$
of cardinality $|V|\leq d$ it holds that
\[
 \Pr\left[\forall x\in V,\;  \|Sx\|^2_2 \in (1\pm\epsilon) \|x\|^2_2 \right]
  \geq 1-\delta.
\]
\end{definition}
An $(\epsilon,\delta,d)$-JLT can be constructed with
$t=O(\epsilon^{-2}\log(d/\delta))$ rows,
which is optimal {(see \cite{kane2011almost} or \cite{jayram2013optimal})}.
While using independent $N(0,1/t)$ Gaussians entries works,
there is a construction with only $O(\eps^{-1}\log(1/\delta))$
non-zero entries per column~\cite{kane2014sparser}.

The case $p=4$ has a particularly short and simple analysis,
whenever $G_1$ and $G_2$ are independent $(\eps,\delta,n)$-JLT matrices,
which we can achieve with $t=O(\eps^{-2}\log n)$.
The first idea is to ``peel off'' $G_i$ from both sides,
using that for any PSD matrix $M$,
with high probability $\tr(G_iMG_i^T)\in(1\pm\eps)\tr(M)$
(see Lemma~\ref{lemma:jlt preserves trace} for a precise statement).
A second idea is to use the identity $\tr(BC) =\tr(CB)$
to rewrite $\tr(AA^TG_2^TG_2AA^T) = \tr(G_2AA^TAA^TG_2^T)$.
Now using the first idea once again, we are likely to arrive at an
approximation to $\tr(AA^T$ $AA^T)=\norm{A}_{S_4}$.
The full details are given in Section~\ref{sec:S4}.

The sketching method extends from $p=4$ to any integer $p\ge 2$,
but the simple analysis above breaks
(because for $p>4$ the ``inside'' matrix $M$ is no longer PSD)
and thus our analysis is much more involved.
We first analyze $G_i$'s with independent Gaussian entries,
by a careful expansion of the fourth moment of $X$,
which exploits certain cancellations occurring (only) for Gaussians.
We then consider $G_i$'s that are sampled from a particular sparse JLT
due to \cite{thorup2004tabulation},
and employ a symmetrization-and-decoupling argument to compare
the variance of $X$ in this case with that of Gaussian $G_i$'s.

We make two technical remarks.
First, proving $\EX[X] = \tr(A^p)$ is straightforward.
Indeed, by the second idea above, we can rewrite
$X
 = \tr(G_1 AG_2^TG_2 AG_3^T\cdots G_pA G_1^T)$
as $X = \tr(G_1^T G_1 AG_2^TG_2A\cdots$ $G_p^TG_p A)$.
Now using $\EX[G_i^TG_i] = I$ together with linearity
of trace and of expectation, we obtain that $\EX[X] = \tr(A^p)$.
Second, after setting $t=O(n^{1-2/p})$ (independent of $\eps$),
our bound on the variance is $O(\EX[X]^2)$,
which we can decrease in a standard way, taking $O(1/\eps^2)$ repetitions.
See Sections~\ref{sec:SpGaussian} and~\ref{sec:faster} for details.

The multi-pass streaming algorithm is implemented slightly differently,
in that $G_1\in\R^{1\times n}$, i.e., has only one row.
The other matrices $G_2,\dots,G_p\in\R^{t\times n}$ are as before,
although we now set $t=O(n^{1-1/(p-1)})$.
Our estimator $X$ 
can be computed in $\ceil{p/2}$ passes with space only $2t$ as follows.
In the first pass, compute vectors
$X_L \gets G_1AG_2^T\in\R^{1\times t}$
and $X_R\gets G_p^TAG_1\in\R^{t\times 1}$,
and then on the $i$-th pass update
$X_L\gets X_LG_i^TAG_{i+1}$
and $X_R\gets G_{p-i+1}AG_{p-i+2}^TX_R$.
Notice that the computation in each pass is linear in $A$.
For even $p$, after completing $p/2$ passes,
compute and output $X'=X_LX_R\in\R$ (and similarly for odd $p$).
This $X'$ is similar to the estimator $X$ described above,
except for the new dimensions of the $G_i$'s.
See Sections~\ref{sec:multipass} and~\ref{sec:faster}.

This multi-pass algorithm offers a very significant space savings over
the one-pass algorithm.
It is also a bit surprising because it is getting close to
the corresponding vector norm, namely, $\ell_p$-norm on $\R^n$,
for which the optimal space for $O(p)$ passes is $\tO(n^{1-2/p})$ bits.
In fact, for the vector norm, $O(p)$ passes do not significantly
reduce the storage needed compared with one pass,
which stands in sharp contrast to Schatten $p$-norms.
As mentioned before, if $p$ is even then the algorithms extends to arbitrary
$A\in\R^{m\times n}$ by a standard argument.

\subsection{Lower Bound for PSD Matrices}
\label{sec:IntroLB4PSD}
Recent work~\cite{li2016approximating} has improved the storage lower bound
for estimating Schatten $p$-norms for non-integer values of $p$,
by showing that $(1+\epsilon)$-approximation (in the one-pass entry-wise model)
requires storage $n^{1-g(\epsilon)}$, 
for some function $g(\epsilon)\to 0$ as $\epsilon \to 0$,
even for a sparse matrix. 
This contrasts with our algorithms for PSD matrices 
(from Section~\ref{sec:IntroSp}), where the
exponent is independent of $\epsilon$ and bounded away from $1$.
However, the hard distribution used by~\cite{li2016approximating} is not
over PSD matrices, 
leaving open the possibility that PSD matrices admit algorithms 
that use storage $O(n^c)$ for $c<1$ independent of $\epsilon$.

We close this gap in Section~\ref{sec:LB4PSD}, 
by adapting the lower bound of~\cite{li2016approximating} to PSD matrices, 
to show, for every non-integer $p>0$, 
a storage lower bound of $\Omega(n^{1-g'(\epsilon)})$ 
for some function $g'(\epsilon)\to 0$ as $\epsilon \to 0$
(again, in the one-pass entry-wise model and even for a sparse matrix).
A key feature of our lower bounds for PSD matrices is that they hold in the model in which each entry of the matrix occurs exactly once in the stream. This models applications where the matrix resides in external memory and is being streamed through main memory; in such a model multiple updates to an entry may not appear. While it is possible to obtain lower bounds for PSD matrices by embedding the multiplayer \textsc{Set-Disjointness} lower bound \cite{bar2002information} for vectors onto the diagonal of a matrix, to apply such lower bounds the diagonal entries need to be incremented repeatedly, that is, one such diagonal entry needs to be updated $n^{\Omega(1)}$ times. In contrast, in our lower bounds each matrix entry occurs exactly once in the stream, i.e., there are no updates to entries.

\subsection{Results for Row-Order Model}
\label{sec:IntroRowOrder}

For sparse matrices, 
estimating Schatten $p$-norms in the row-order model can be reduced
to estimating Schatten $(p/2)$-norms in the turnstile model. 
Consider estimating $\|A\|_{S_p}^p$ for some sparse matrix $A$. 
The algorithm first forms $A^TA = \sum_i A_i^T A_i$ ``on the fly'',
by reading each row $A_i$ and immediately generating a stream of updates 
that corresponds to the non-zero entries in $A_i^T A_i$, 
and then it can just estimate the Schatten $(p/2)$-norm of that stream, 
because $\|A^TA\|_{S_{p/2}}^{p/2} = \|A\|_{S_p}^p$. 
Observe that each row $A_i$ has only $O(1)$ non-zero entries,
hence also $A_i^T A_i$ has only $O(1)$ non-zero entries,
and the algorithm only needs $O(1)$ space to generate the updates to $A^TA$.
Moreover, since $A$ is sparse, also $A^TA$ is sparse. 
It was shown in \cite{li2016approximating} how to estimate the 
Schatten $p$-norm, for an even integer $p$, 
using $\tilde O_{p,\epsilon}(n^{1-2/p})$ bits of space, even in the turnstile model.
For $p\in 4\Z$, the above yields an algorithm in the row-order model 
that uses $\tilde O_{p,\epsilon}(n^{1-4/p})$ bits of space for sparse matrices. 

In Sections~\ref{sec:RowOrder} and \ref{sec:row_order_ub}, 
we study the problem in the row-order model for all $p>0$. 
When $p$ is not an even integer, 
we prove that $(1+\epsilon)$-approximating the Schatten $p$-norm
in the one-pass entry-wise model  
requires $\Omega_\epsilon(n^{1-g(\epsilon)})$ bits of space
where $g(\epsilon)\to 0$ as $\epsilon\to 0$.
This bound holds even for sparse matrices, in which case it is almost tight. 
When $p\ge4$ is an even integer, 
we prove a lower bound of $\Omega_p(n^{1-4/p})$ bits of space, 
matching up to logarithmic factors the algorithm from above for $p\in 4\Z$. 
For the remaining case $p\equiv 2 \pmod 4$, 
we present an algorithm using $\tilde O_{p,\epsilon}(n^{1-4/(p+2)})$ space, 
leaving a slight polynomial gap from the lower bound of $\Omega_p(n^{1-4/p})$.

\begin{table*}[t]
\begin{center}
\begin{tabular}{|llll|}
  \hline
  \multicolumn{4}{|l|}{Problem: Schatten $p$-norm of a sparse matrix in row-order stream} \\
  \hline
  & which $p>0$ & space & \\
  Algorithms: 
  & all $p$ & $\tilde O(n)$ & trivial (by sparsity), $\epsilon=0$ \\
  & $p\equiv 0 \pmod 4$ & $\tilde O_{p,\epsilon}(n^{1-4/p})$ & Section~\ref{sec:IntroRowOrder} \\
  & $p\equiv 2 \pmod 4$ & $\tilde O_{p,\epsilon}(n^{1-4/(p+2)})$ & Theorem~\ref{thm:row_order_ub}, $p\ge 6$  \\
  Lower Bounds:
  & $p\in 2\Z, p\ge 4$ & $\Omega(n^{1-4/p})$ & Theorem~\ref{thm:even-int-lb}, for $\epsilon<\epsilon_0(p)$, even multi-pass \\
  & $p\notin 2\Z$ & $\Omega_t(n^{1-1/t})$ & Theorem~\ref{thm:non-int-lb}, for $\epsilon<\epsilon_0(t,p)$ \\
  \hline
\end{tabular}
\end{center}
\caption{Bounds for $(1+\eps)$-approximation of the Schatten $p$-norm 
of a sparse matrix $A\in\R^{n\times n}$ in the one-pass row-order model. 
Space is counted in bits.
}
\label{tab:RowOrder}
\end{table*}

\subsection{Previous Work}
\label{sec:prev}

The aforementioned algorithm of \cite{li2014sketching}
uses a single sketching matrix $G$,
for example, if $A$ is PSD, then their sketch is $S=GAG^T$,
where $G\in\bbR^{t\times n}$ is a Gaussian matrix.
Its estimate for $\norm{A}_{S_p}$ is produced by
summing over all ``cycles'' $S_{i_1, i_2}S_{i_2, i_3} \cdots S_{i_p, i_1}$,
where $i_1, \dots, i_p\in[t]$ are distinct.
Our sketch improves upon theirs in both update time and query time.
The only other streaming algorithm for Schatten $p$-norm that we are aware
of is that of \cite[Theorem 7]{li2016approximating},
which uses space $O(n^{1-\frac{2}{p}}\poly(\frac{1}{\epsilon}, \log n))$
but works only for matrices that have $O(1)$-entries per row and per column.

One possible approach to improve the update time would be to replace the
Gaussian matrices in  \cite{li2014sketching} with a distribution over
matrices that admit a fast multiplication algorithm.  The analysis done in
\cite{li2014sketching} relies on the Gaussian entries (rotational
invariance, in particular), so the replacement matrix should preserve the
distribution of the sketch.  Kapralov, Potluru, and Woodruff
\cite{kapralov2016fake} present just such a distribution on matrices
$\tilde{G}$, where the multiplication $\tilde{G}A$ can be computed quickly
and $\tilde{G}A$ is close to $GA$ in total variation distance.
Unfortunately, under the distribution of \cite{kapralov2016fake}, or any
other with a similar guarantee on total variation distance, each coordinate
update to $A$ results in a dense rank-one update to the sketch, which means
that the update time is not improved.

Several strong lower bounds are known for approximating Schatten $p$-norms 
and other matrix functions,
both for the dimension of a sketch and for storage requirement (bits).
Li, Nguyen and Woodruff \cite{li2014sketching} prove that for $0\le p<2$
every linear sketch that can approximate rank and Schatten $p$-norm
must have dimension $\Omega(\sqrt{n})$ and every bilinear sketch must have
dimension $\Omega(n^{1-\epsilon})$.
Li and Woodruff \cite{li2016tight} show that every linear sketch
for Schatten $p$-norm, $p\ge 2$, requires dimension $\Omega(n^{2-4/p})$.
In~\cite{li2016approximating}, they prove space complexity lower bounds
that hold even when the input matrix is sparse.
Specifically, they show that one-pass streaming algorithms
which $(1\pm \epsilon)$-approximate various functions of the singular
values,
including Schatten $p$-norms when $p$ is not an even integer,
require $\Omega(n^{1-g(\eps)})$ bits of space
for some function $g(\epsilon)\rightarrow 0$ as $\epsilon\rightarrow 0$.
Additional space lower bounds, e.g., for $p\in[1,2)$,
can be deduced from a general statement of~\cite{AKR15},
see~\cite[Table 1]{li2016approximating}.


%% file: prelim.tex
\section{Notation and Preliminaries}


The space bounds of sketching algorithms in the turnstile model are stated in terms of sketch dimension (number of entries).  
The number of bits required can be larger by a $\log nM$ factor, where $M$ is the absolute ratio of the largest element in the matrix to the smallest.
 We call a matrix a {\em Gaussian matrix} if its entries are independent $N(0,1)$ random variables. 
 A matrix $G$ of dimension $t\times n$ is a {\em column-normalized} Gaussian matrix if $G=G'/\sqrt{t}$, where $G'$ is a Gaussian matrix. 
 Now-standard techniques such as Nisan's Pseudo-random generator or $k$-wise independence can be used to derandomize Gaussian matrices for use in sketching algorithms.
Column-normalized Gaussian matrices serve as JLTs. In particular, there exists a constant $c$ such that if $G$ be a $t\times n$ column-normalized Gaussian matrix with $t\ge \frac{c}{\epsilon^2}\log\frac{d}{\delta}$, then $G$ is a $(\epsilon, \delta, d)$-JLT~\cite{indyk1998approximate}.


%% file: sec_schatten.tex
\section{New Estimator for PSD Matrices (and Integer \mathp) \label{sec:Sp}}
\label{sec:Sp}

\newcommand{\compsq}{
\begin{tikzpicture}[baseline=-0.65ex,scale=0.5]
\draw (-1, 1) -- (-1, -1) -- (1, -1) -- (1, 1) -- cycle;
\draw (1, 1) -- (-1, -1);
\draw (1,-1) -- (-1, 1);
\end{tikzpicture}
}
\newcommand{\leftrightsq}{
\begin{tikzpicture}[baseline=-0.65ex,scale=0.5]
\draw (-1, 1) -- (-1, -1);
\draw (1, 1) -- (1, -1);
\end{tikzpicture}
}
\newcommand{\updownsq}{
\begin{tikzpicture}[baseline=-0.65ex,scale=0.5]
\draw (-1, 1) -- (1, 1);
\draw (-1, -1) -- (1, -1);
\end{tikzpicture}
}
\newcommand{\crosssq}{
\begin{tikzpicture}[baseline=-0.65ex,scale=0.5]
\draw (-1, 1) -- (1, -1);
\draw (-1, -1) -- (1, 1);
\end{tikzpicture}
}
\newcommand{\alldots}{
\begin{tikzpicture}[baseline=-0.65ex,scale=0.5]
\filldraw[black] (1, 1) circle (1pt);
\filldraw[black] (1, -1) circle (1pt);
\filldraw[black] (-1, -1) circle (1pt);
\filldraw[black] (-1, 1) circle (1pt);
\end{tikzpicture}
}
\newcommand{\singleVertlineA}{
\begin{tikzpicture}[baseline=-0.65ex,scale=0.5]
\filldraw[black] (-1, 1) circle (1pt);
\filldraw[black] (-1, -1) circle (1pt);
\draw (1,-1) -- (1, 1);
\end{tikzpicture}
}
\newcommand{\singleVertlineB}{
\begin{tikzpicture}[baseline=-0.65ex,scale=0.5]
\filldraw[black] (1, 1) circle (1pt);
\filldraw[black] (1, -1) circle (1pt);
\draw (-1,-1) -- (-1, 1);
\end{tikzpicture}
}
\newcommand{\singleCrossA}{
\begin{tikzpicture}[baseline=-0.65ex,scale=0.5]
\filldraw[black] (1, 1) circle (1pt);
\filldraw[black] (-1, -1) circle (1pt);
\draw (1,-1) -- (-1, 1);
\end{tikzpicture}
}
\newcommand{\singleCrossB}{
\begin{tikzpicture}[baseline=-0.65ex,scale=0.5]
\filldraw[black] (-1, 1) circle (1pt);
\filldraw[black] (1, -1) circle (1pt);
\draw (1,1) -- (-1, -1);
\end{tikzpicture}
}
\newcommand{\singleTriangleA}{
\begin{tikzpicture}[baseline=-0.65ex,scale=0.5]
\filldraw[black] (1, -1) circle (1pt);
\draw (1,1) -- (-1, -1) -- (-1, 1) -- cycle;
\end{tikzpicture}
}

\newcommand{\singleTriangleB}{
\begin{tikzpicture}[baseline=-0.65ex,scale=0.5]
\filldraw[black] (-1, 1) circle (1pt);
\draw (-1,-1) -- (1, -1) -- (1, 1) -- cycle;
\end{tikzpicture}
}

The main result in this section is a new one-pass streaming algorithm 
for estimating the Schatten $p$-norm, for integer $p\ge 2$.
When $p$ is odd, it additionally requires that the input matrix is PSD.
The first version of this algorithm, 
described in Section~\ref{sec:SpGaussian},
has the same storage requirement of $\tO_p(n^{2-4/p}/\epsilon^{2})$ bits
as the previous algorithm of \cite{li2014sketching} that uses cycle sums,
but has simpler analysis and faster query time\footnote{In~\cite{kong2016spectrum}, Kong and Valiant independently improve the algorithm in~\cite{li2014sketching} to the same runtime as Theorem~\ref{thm:Sp} in this paper by considering only ``increasing cycles''.}, 
which is roughly matrix multiplication time, $n^\omega$, instead of $n^p$.
Moreover, it is based on a new method that leads to a $\ceil{p/2}$-pass algorithm with storage requirement $\tO_p(n^{1-1/(p-1)}/\eps^2)$ bits,
as described in Section~\ref{sec:multipass}.
Previously, the algorithm in \cite[Theorem 6.1]{woodruff2014sketching}
has the same number of passes but larger storage requirement $O(n/\eps^{2})$.%
\footnote{We note that also in \cite[Theorem 6.1]{woodruff2014sketching}
it is required that $p$ is even or that the input matrix is PSD, 
but this is erroneously omitted.
}
Finally, we improve the update time, as described in Section~\ref{sec:faster},
by employing the sketching matrices $G_i$ 
that are certain sparse matrices instead of Gaussians.

We start in Section~\ref{sec:S4} with the case $p=4$, 
which is based on the same sketch but is significantly easier to analyse.

\subsection{Schatten $4$-Norm using JLT matrices}
\label{sec:S4}

\begin{theorem}\label{thm:S4}
Let $G_1, G_2\in\R^{t\times n}$ be independent $(\epsilon, \frac{\delta}{n}, 1)$-JLT matrices. 
Then for every $A\in\R^{n\times m}$, 
\[
  \Pr\Big[  
    \tr(G_1AA^TG_2^TG_2AA^TG_1^T) \in (1\pm 2\epsilon)^2 \norm{A}_{S_4}^4
  \Big]
  = 1-2\delta.
\]
Thus, one can find a $(1\pm\epsilon)$-approximation to the Schatten-$4$ norm of a general matrix $A\in\R^{n\times m}$ using 
a linear sketch of dimension $O(\epsilon^{-2}n\log n)$.
\end{theorem}
Before proving the theorem, we remark that if each column of $G_i$ has only $s$ non-zero entries, it is easy to see that 
the update time of this linear sketch is $O(s)$,
assuming any entry of $G_1$ and $G_2$ can be accessed in $O(1)$ time
(in a streaming algorithm, the entries are usually computed
from a small random seed in $\polylog(n)$ time). 
The query time is dominated by multiplying a matrix
of size $t\times n$ with one of size $n\times t$, 
and thus take $O(t^\omega\cdot n/t)=\tO(n^\omega/\eps^{2(\omega-1)})$.

Now we prove Theorem~\ref{thm:S4}, for which we need the following lemma.
\begin{lemma} \label{lemma:jlt preserves trace}
Let $G\in\R^{t\times n}$ be an $(\epsilon, \delta/n, 1)$-JLT matrix.
Then for every PSD matrix $A\in\R^{n\times n}$,
\[
  \Pr\Big[  \tr(GAG^T) \in (1\pm\epsilon) \tr(A) \Big] \ge 1-\delta.
\] 
\end{lemma}
\begin{proof}
By the Spectral Theorem, $A=U\Lambda U^T$, 
where $\Lambda$ is a diagonal matrix and $U$ is an orthonormal matrix. 
Then $G'=GU$ is a still $(\epsilon, \delta/n, 1)$-JLT. 
Thus 
\begin{multline*}
  \tr(GAG^T) 
  = \tr(G'\Lambda G'^T) 
  = \tr(\sqrt{\Lambda}G'^TG'\sqrt{\Lambda}) 
  = \sum_{i=1}^n \lambda_i e_i^TG'^TG'e_i
  = \sum_{i=1}^n \lambda_i \norm{G'e_i}_2^2.
\end{multline*}
By the JLT guarantee and a union bound, with probability at least $1-\delta$, 
for all $i\in [n]$ we have $\norm{G'e_i}_2^2 \in[1-\epsilon, 1+\epsilon]$,
in which case $\tr(GAG^T) \in (1\pm\epsilon) \tr(A)$.
\end{proof}

\begin{proof}[of Theorem~\ref{thm:S4}]
Apply Lemma~\ref{lemma:jlt preserves trace} to the PSD matrix $AA^TAA^T$,
to get that with probability at least $1-\delta$ (over the choice of $G_2$), 
\begin{align*}
  \tr(G_2AA^TAA^TG_2^T) 
  \in(1\pm2\eps) \tr(AA^TAA^T)
  =  (1\pm2\eps) \norm{A}_{S_4},
\end{align*}
where the left-hand side is equal to $\tr(AA^TG_2^TG_2AA^T)$,
by the identity $\tr(MM^T) = \tr(M^TM)$.
Now suppose (by conditioning) that $G_2$ is already fixed,
and apply the same lemma to the PSD matrix $AA^TG_2^TG_2AA^T$,
to get that with probability at least $1-\delta$ (over the choice of $G_1$), 
\[
  \tr(G_1AA^TG_2^TG_2AA^TG_1^T) 
  \in (1\pm 2\epsilon)\tr(AA^TG_2^TG_2AA^T).
\]
The proof follows by a union bound.

The linear sketch of $A$ consists of the two matrices $G_1 A$ and $G_2 A$,
which suffices to estimate $\norm{A}_{S_4}^4$ as above with $\delta=1/8$.
This sketch is linear and its dimension is $2tn$, 
where we can use say Gaussians to obtain $t=O(\eps^{-2}\log n)$.
\end{proof}

\subsection{Schatten $p$-norm Using Gaussians}
\label{sec:SpGaussian}

We now design a sketch for Schatten-$p$ norm that uses column-normalized Gaussian matrices. 
We will later extend and refine it to improve the per-update processing time.

\begin{theorem} \label{thm:Sp}
For every $0<\eps<1/2$ and integer $p\ge 2$,
there is an algorithm that outputs at $(1\pm\epsilon)$-approximation to
the Schatten-$p$ norm of a PSD matrix $A\in\R^{n\times n}$
using a randomized linear sketch of dimension $s = O_p(\epsilon^{-2} n^{2-4/p})$.
The update time (for each entry in $A$) is $O(s)$ 
and the query time (for computing the estimate) is $O(\eps^{-2}n^{(1-2/p)\omega})$, 
where $\omega< 2.373$ is the matrix multiplication constant.

If $p$ is even, the above algorithm extends to a general matrix 
$A\in \R^{n\times m}$.

\end{theorem}

The first part of the theorem (for PSD matrices) follows directly 
from Proposition~\ref{prop:gaussian sketch for sp integer} below.
The proposition is applicable to all symmetric matrices, but $\|A\|^p_{S_p}=\tr(A^p)$ only for PSD matrices or even $p$.
The linear sketch stores $G_iAG_{i+1}^T$ for $i=1,\ldots,p$, 
where by convention $G_{p+1}=G_1$,
repeated independently in parallel $O_p(1/\eps^2)$ times.
Thus, the sketch has dimension $O_p(\eps^{-2} t^2)$.
The estimator is obtained by computing the $O_p(1/\eps^2)$ 
independent copies of $X$ and reporting their average.
To analyze its accuracy,
notice that a PSD matrix $A$ satisfies $\EX[X]=\tr(A^p)=\norm{A}_p^p$. 
Then setting $t = n^{1-2/p}$ gives $\Var(X) \leq O_p(\norm{A}_{S_p})^{2p}$
and averaging multiple independent copies of $X$ reduces the variance.

The second part (for general matrices), 
follows by using the same sketch for the symmetric 
matrix
$
B=\left(\begin{smallmatrix}
0 & A\\
A^T & 0
\end{smallmatrix}\right),
$ 
because the nonzero singular values of $B$ are those of $A$ repeated twice
and $\norm{B}_{S_p}^p=2 \norm{A}_{S_p}^p = 2 \tr(A^p)$,
where the last equality uses the assumption that $p$ is even.

Because the correctness of the algorithm comes by bounding the variance of $X$, it is enough that the entries in each Gaussian matrix are four-wise independent, which is crucial for applications with limited storage like streaming. 

\begin{proposition}
\label{prop:gaussian sketch for sp integer}
\label{PROP:GAUSSIAN SKETCH FOR SP INTEGER}
For integer $p\ge 2$ and $t\ge 1$, let $G_1, \dots, G_{p}$ 
be independent $t\times n$ column-normalized Gaussian matrices.
Then for every symmetric matrix $A\in\R^{n\times n}$, 
the estimator $X=\tr\big(G_1 AG_2^TG_2 A\ldots G_p^T$ $G_pAG_1^T\big)$ satisfies
\begin{align*}
  \EX[X] = \tr(A^p)
\qquad\text{and}\qquad
  \Var(X)
  = 
  O_p\!\left(1\!+\!\sum_{z=2}^{\lfloor \frac{p}{2}\rfloor + 1}\!\left(\frac{n^{1-\frac{2}{p}}}{t}\right)^{\!\!z}
    + \sum_{z=2}^{p}\left(\frac{n^{1-\frac{2}{z}}}{t}\right)^{\!\!z}
          \right)\norm{A}_{S_p}^{2p}.
\end{align*}
\end{proposition}
The full proof of this proposition appears in postponed to Section~\ref{sec:proof gaussian sketch for sp integer}. We outline the general idea here. It is standard that a Gaussian 
matrix is rotational invariant, i.e., $G$ and $GU$ are identically
distributed for any orthogonal matrix $U$. Thus, by the Spectral
Theorem, instead of considering symmetric matrix $A = U\Lambda U^T$, 
we can consider only its diagonalization $\Lambda$. 

The proof of this proposition proceeds first by 
expanding $X$ in terms of inner products
of columns of the matrix $G$, i.e.,
$X
  = \sum_{i_1, i_2, \ldots, i_p\in[n]} 
    \lambda_{i_1}\lambda_{i_2}\ldots \lambda_{i_p}
    \cdot $ $
    \langle g^{(1)}_{i_1}, g^{(1)}_{i_2}\rangle \cdot$ $
    \langle g^{(2)}_{i_2}, g^{(2)}_{i_3}\rangle
    \ldots \langle g^{(p)}_{i_p}, g^{(p)}_{i_1}\rangle
$,
where $\lambda_i$ is the $i$-th eigenvalue of $A$ and $g_{i_j}^{(j)}$ is the $i_j$-th column of $G_j$. 
We then expand $\EX(X^2)$. 
The non-zero terms in $\EX(X^2)$ are
composed by only those terms of even powers in every eigenvalue. 
Computing the expectation of each term is straightforward because the entries of $G$ are independent Gaussian random variables, but the crux of the proof is in bounding the sum of the terms.
We introduce a collection of diagrams that aid in enumerating the terms according to their structure and computing the sum.

\subsection{Multi-Pass Algorithm}
\label{sec:multipass}

The proof of Proposition~\ref{prop:gaussian sketch for sp integer} 
relies on the matrices $G_i$ being Gaussians in two places.
First, we  assume that the matrix $A$ is diagonal,
and in general we need to consider $G_iU$ instead of $G_i$.
Second, the columns of these matrices have small variance/moments,
as described in \eqref{eqn:sqr_exp1}-\eqref{eqn:sqr_exp2}.
We now generalize the proof to relax these requirements 
(e.g., to $4$-wise independence) and obtain a multi-pass algorithm. 

\begin{lemma}
\label{cor:reduced size sketch for s norm}
\label{COR:REDUCED SIZE SKETCH FOR S NORM}
For integers $p\ge 2$ and $1\le t'\le t$,
let $G_1\in\R^{t' \times n}$ and $G_2, \ldots, G_{p}\in\R^{t \times n}$
be independent column-normalized Gaussian matrices
with $4$-wise independent entries.
The for every symmetric matrix $A\in\R^{n\times n}$,
the estimator $X=\tr\left(G_1AG_2^TG_2A \ldots G_p^TG_pAG_1^T\right)$ satisfies
\begin{align*}
  \EX[X]=\tr(A^p)
\qquad\text{and}\qquad
  \Var(X)
  = 
 O_p\bigg(1+
      \sum_{z=2}^{\lfloor p/2\rfloor} \frac{n^{z-1-2(z-1)/p}}{t't^{z-1}}
      + \sum_{z=2}^{p} \frac{n^{z-2}}{t't^{z-1}}
    \bigg) \|A\|_{S_p}^{2p}.
\end{align*}
\end{lemma}
The proof of this lemma is postponed to~\ref{sec:proof reduced size sketch for s norm}. 
It is a direct corollary of the proof of Proposition~\ref{prop:gaussian sketch for sp integer},
 except that $t'$, the size of the first sketch matrix, is emphasized.

We can now use the above sketch to approximate the Schatten $p$-norm using $\tilde{O}(n^{1-1/(p-1)})$ bits of space with $\lceil p/2\rceil$ passes over the input. 

\begin{theorem} \label{thm:multipass}
Let $p\ge 2$ be an even integer. 
There is a $\ceil{p/2}$-pass streaming algorithm, 
that on input matrix $A\in\R^{n\times m}$ with $n\ge m$ given as a stream,
outputs an estimate $X$ such that with probability at least $0.9$,
$  X \in (1\pm\eps) \norm{A}_{S_p}^p 
$
and uses $O_p(n^{1-1/(p-1)}/\epsilon^2)$ words of space. 
The above extends to all integers $p\ge 2$ if $A$ is PSD.
\end{theorem}

\begin{proof}[of Theorem~\ref{thm:multipass}]
Without loss of generality we may assume that $A$ is symmetric
as argued in the proof of Theorem~\ref{thm:Sp}.
We first describe a basic algorithm that produces an estimator 
for $\|A\|_{S_p}^p$ that is unbiased and has variance $O_p(\|A\|_{S_p}^{2p})$.
We will later decrease the variance to $O(\epsilon^2\|A\|_{S_p}^{2p})$ 
using the standard technique of independent repetitions in parallel.

The basic algorithm uses a pseudo-random generator
to produce a four-wise independent column-normalized Gaussian matrix.
In fact, it samples $p$ such matrices, namely,
$G_1\in\R^{1\times n}$ and $G_2,\dots,G_p\in\R^{t\times n}$ for $t=O(n^{1-1/(p-1)})$,
where the $p$ matrices are independent of each other.
In the first pass, the algorithm computes $G_1 AG_2^T$ and $G_p A G_1^T$,
and stores them in memory. 
Notice that these are linear sketches of $A$, each dimension $t$. 
In the second pass, the algorithm uses these results to compute
$(G_1 AG_2^T) G_2AG_3^T$ and $G_{p-1} A G_p^T(G_p A G_1^T)$
which are again linear sketches of the stream $A$ 
(given the result of the first pass), each of dimension $t$. 
Continuing in this manner until pass number $\lceil p/2\rceil$,
the algorithm stores in memory the vectors 
$h=  G_{\lfloor p/2\rfloor} A G_{\lfloor p/2\rfloor + 1}^T \cdot G_p A G_1^T$ and 
$h'^T =G_1 AG_2^T \cdots G_{\lfloor p/2\rfloor-1} A G_{\lfloor p/2\rfloor}^T$,
each of dimension $t$.
Now compute $Y=h'^Th$. 
By Lemma~\ref{cor:reduced size sketch for s norm} we have that 
$\EX[Y] = \tr(A^p) = \sum_i \lambda_i^p$
(where in the case that $p$ is odd we use the assumption that $A$ is PSD).
Thus, $Y$ is an unbiased estimator for $\|A\|_{S_p}^p$,
and it remains to bound its variance.
By Lemma~\ref{cor:reduced size sketch for s norm},
\begin{align*}
 \Var(Y)
  = 
  O_p \bigg(
    \sum_{z=2}^{\lfloor p/2\rfloor} \frac{n^{z-1-2\frac{z-1}{p}}}{n^{z-1 -\frac{z-1}{p-1}}}
    + \sum_{z=2}^{p} \frac{n^{z-2}}{n^{z-1 - \frac{z-1}{p-1}}}
    \bigg) \|A\|_{S_p}^{2p}
  = O_p(\|A\|_{S_p}^{2p}).
\end{align*}

By repeating the basic algorithm $O_p(1/\epsilon^2)$ times in parallel
and reporting the average of their estimates $Y$, 
we obtain estimator $X$ for $\|A\|_{S_p}^p$ that is unbiased 
and has variance at most $\tfrac19\epsilon^2 \|A\|_{S_p}^{2p}$. 
The correctness of this estimator follows by Chebyshev's ineuqality.
The basic algorithm is required to store 
$2p$ intermediate vectors of dimension $t$ 
and random seeds for the $p$ Gaussian matrices. 
By standard techniques, 
the length of the seeds is $O_p(\polylog n)$ bits.
The final algorithm stores these for all the $O_p(1/\epsilon^2)$ repetitions,
and Theorem~\ref{thm:multipass} follows.
\end{proof}

\subsection{Faster Update Time}
\label{sec:faster}
Since Gaussian matrices are dense, a change to one coordinate of the input matrix $A$ may lead to a change of every entry in the sketch.
This means long update times for a streaming algorithm based on the sketch. 
In this section we extend our result for Gaussian sketching matrices to a distribution over $\{-1,0,1\}$ valued matrices with only one non-zero entry per column. 
The new sketch can be used to improve the update time of algorithms in the last two sections.

\begin{definition}[Sparse \zdsketch] 
Let $\calD_{t, n}$ be the distribution over matrices $G :=Z D\in\R^{t\times n}$,
where $Z=(z_1, z_2, \ldots, z_n)\in\R^{t\times n}$ and
$D=\diag(d_1, d_2, \ldots, d_n)$ are generated as follows.
Let $h:[n]\rightarrow [t]$ be a $4$-wise independent hash function,
and set $Z_{i,j} = \indice{i=h(j)}$,
i.e., in each $z_j$ only the $h(j)$-th coordinate is set to $1$, 
and all other coordinates are $0$.
The diagonal entries of $D$ are four-wise independent uniform $\{-1, 1\}$ random variables, and $D$ is independent from $Z$. 
\end{definition}
Notice that each column of $G$ has a single non-zero entry, 
which is actually a random sign, 
and the $n$ columns are four-wise independent.
This random matrix $G$ is similar to the sketching matrix used 
in \cite{thorup2004tabulation} to speed up the update time 
when estimating the second frequency moment of a vector in $\R^n$.
{Also note that the $ZD$-sketch is a version of sparse JL matrices (see e.g., \cite{kane2014sparser, dasgupta2010sparse}). In this paper we do not aim at optimizing the sparsity as we focus on approximating Schatten norms.
}

It is fairly easy to show that 
\zdsketch~works for approximating Schatten $p$-norm of matrices
 with all entries non-negative. The proof is presented in Section~\ref{sec:simple proof for matrices with all positive entries}.
We now show that the conclusion of Theorem~\ref{thm:Sp} 
and Theorem~\ref{thm:multipass} still hold if we replace the Gaussian matrices in the sketch with independent samples from the sparse \zdsketch. 
A major difficulty that arises in replacing the Gaussian matrix 
with the sparse \zdsketch~is the latter's lack of rotational invariance. 
To prove Theorem~\ref{thm:Sp} we were able to expand $X^2$ in terms of the eigenvalues of $A$ and compute the expectation term-by-term, but this is not possible for the sparse \zdsketch.
For example,
let $G$ be a Gaussian matrix, for any orthogonal matrix $U$, the matrix $GU$ is again a Gaussian matrix with an identical distribution to $G$. This does not hold for sparse \zdsketch. As a consequence, 
in the expansion of $\EX(X^2)$ in the proof of Proposition~\ref{prop:gaussian sketch for sp integer}, the non-zero terms
would also include those monomials of odd powers of $\lambda_i(A)$. For example, for Schatten $3$-norm, one cannot bound $\sum_{i_1, i_2, \ldots i_6\in [n]} \prod_{j=1}^6 \lambda_{i_j}$ by $O(\|A\|_{S_3}^{6})$. But this term appears 
in the expansion of $\EX(X^2)$ of the Schatten $3$-norm estimator
if using the sparse \zdsketch~matrices.

To resolve this problem, we use a technique similar to the proof of 
Hanson-Wright Inequality in \cite{rudelson2013hanson} to bound the variance of $X$.
The proof is composed of three major steps. The first step is to decouple the dependent summands by injecting independence. 
The second step is to replace the independent random vectors with fully independent
Gaussian vectors while preserving the variance. We can then apply our techniques for Gaussians to bound the variance of the final random variable. 
The case $p=1$ is useful to illustrate the technique, even though Schatten 1-norm approximation can be easily accomplished in other ways. 
Let $G\in\R^{t\times n}$ be the sparse JLT matrix and let $A\in\R^{n\times n}$ be PSD.
The sketch is $GAG^T$ and 
\begin{equation}\label{eq: sparse JLT sum expansion p=1}
\tr(GAG^T)-\tr(A) = \sum_{i\neq  j}a_{i,j}\langle g_{i}, g_j\rangle.
\end{equation} Since $i\neq j$, 
$g_i$ and $g_j$ are independent. However the summands are subtly dependent. We first decouple
the summand by choosing $\delta_i\sim $Bernoulli$(1/2)$,  and write $\langle g_i, g_j\rangle = 4\EX(\delta_i(1-\delta_j)\langle g_i, g_j\rangle)$. Let $V=\{i: \delta_i=1\}$, then $\sum_{i\neq j}$ $a_{i, j}\langle g_i, g_j\rangle=4\EX_\delta \sum_{i\in V, j\in \bar{V}}a_{i, j}\langle g_i, g_j\rangle$.  {Thus conditioning on $\delta$ and $\{g_j: j\in \bar{V}\}$, the set 
$\{\langle g_i, \sum_{j\in \bar{V}}a_{i, j}g_j\rangle: i\in V\}$ is a set of independent random variables}. We can match 
these random variables with Gaussian random variables of the same variance, and thus replace $g_i$
with independent Gaussian vectors. The same process can be repeated for $g_j: j\in \bar{V}$, and replace
every vector $g_i: i\in[n]$ by independent Gaussian vectors. 
This lets us apply similar techniques as used in the proof of Proposition 
\ref{prop:gaussian sketch for sp integer} to bound the variance of the resulting random variable, and thus 
bound the variance of the original random variable $\tr(GAG^T) - \tr(A)$.

The analogue of \eqref{eq: sparse JLT sum expansion p=1} for the case of our general estimator, $X-\tr(A^p)$,
is much more complicated than the $p=1$ case.
We observe that {these terms can be grouped
as a sum of products of consecutive \emph{walks}, i.e.,\\
$a_{i_1,i_2}a_{i_2, i_3}\ldots a_{i_{z}, j_{z+1}}\langle
 g_{j_{z+1}}^{(z+1)}, g_{i_{z+1}}^{(z+1)}
\rangle$ for some $z$. Notice that $\langle
 g_{j'}^{(z')}, g_{j'}^{(z')}
\rangle = 1$ for any $j'$ and $z'$}. For each walk, we can apply similar idea to replace 
the $g_i$ vectors with independent Gaussian vectors. Again,
we apply similar techniques as used in the proof of Proposition 
\ref{prop:gaussian sketch for sp integer} to bound the variance
of each group. As a result, when replacing the Gaussian matrices 
by sparse JLT matrices, Lemma~\ref{cor:reduced size sketch for s norm}
still holds. 

Using the sparse \zdsketch, we are able to achieve the same space bound and query time as in 
Theorem~\ref{thm:Sp} and Theorem~\ref{thm:multipass}. But our update time
is improved to $O(1/\epsilon^2)$. We present the full statement of our theorem below. The full proof 
can be found in are presented in Section \ref{sec:proof for general matrices}. 

\begin{theorem}
\label{thm:fast update algorithm general}
For every $0<\eps<1/2$ and integer $p\ge 2$,
there is a randomized one-pass streaming algorithm $\calA$ 
with space requirement $O(n^{2-4/p}/\epsilon^2)$, 
that given as input a PSD matrix $A\in \bbR^{n\times n}$, 
outputs with high probability a $(1+\eps)$-approximation of $\norm{A}_{S_p}^p$.
The algorithm processes an update in time $O(1/\epsilon^2)$,
and computes the output (after the updates)
in time $O(n^{(1-2/p)\omega})/\epsilon^2$,
where $\omega<3$ is the matrix multiplication constant. 

There is similarly a randomized $\lceil p/2\rceil$-pass streaming algorithm $\calB$ with space requirement \\$O(n^{1-1/(p-1)} /\epsilon^2)$, 
update time in a pass $O(1/\epsilon^2)$, 
and output time $O(n^{(1-2/p)}/\epsilon^2)$.

For even $p\ge 2$,
both algorithms extend to general input $A\in\bbR^{n\times m}$ with $m\le n$.
\end{theorem}


%% file: row_order.tex
\section{Lower Bound For PSD Matrices}
\label{sec:LB4PSD}

\begin{theorem}\label{thm:PSD_lb}
Suppose that $p>0$ and $X\in \R^{n\times n}$ is a PSD matrix 
given in the entry-wise streaming model.
\begin{enumerate}[(a)]\compactify
\item When $p\in\Z$, there is $c=c(p)>0$ such that every one-pass streaming algorithm that $(1+c)$-approximates $\|X\|_{S_p}$ with probability $2/3$
must use $\Omega_p(n^{1-2/p})$ bits of space for even $p$, 
and $\Omega_p(n^{1-2/(p-1)})$ bits of space for odd $p$. 
\item When $p\not\in\Z$, for every integer $t\geq 2$, 
there is $c = c(p, t) > 0$ such that every one-pass streaming algorithm that $(1+c)$-approximates $\|X\|_{S_p}$ with probability $2/3$ 
must use $\Omega_{p,t}(n^{1-1/t})$ bits of space.
\end{enumerate}
\end{theorem}

\begin{proof}
Let $M$ be drawn from the hard input distribution for even integer $p$ in~\cite{li2016approximating}, which involves an integer parameter $t$ but does not depend on the value of $p$. This $M$ is drawn from one of two distributions with the properties that (i) for each even integer $r\geq 2t$, there exist a threshold $L$ and a small constant $\eta$ which both depend on $r$ and $t$ such that with high probability, $\|M\|_r^r\geq (1+\eta)L$ when $M$ is drawn from one distribution and $\|M\|_r^r\leq (1-\eta)L$ when $M$ is drawn from the other distribution; (ii) for any even integer $r < 2t$, there is no such gap in $\|M\|_r^r$ between the two distributions; (iii) distinguishing which distribution $M$ is drawn from requires $\Omega_t(n^{1-1/t})$ bits of space for one-pass streaming algorithms, even in the insertion-only model.

It was also proved in~\cite{li2016approximating} that the maximum singular value of $M$ is at most $t$. Then $A = \left(\begin{smallmatrix}
tI_n & M\\
M^T & tI_n
\end{smallmatrix}\right)$ is positive semidefinite since its eigenvalues are $t\pm\sigma_1(M)$, ..., $t\pm\sigma_n(M)$, all of which are non-negative. 
We shall show that there is a constant-factor gap in $\|A\|_{S_p}^p$ when $M$ is drawn from the two distributions, then the same lower bound in property (iii) above follows.

Consider two distributions over the PSD matrices of the above form, induced by the two distributions of $M$, respectively.
Recall that if $p>0$ is not an integer, 
\[
  \forall |x|\leq 1,
  \quad
  (1+x)^p = \sum_{k=0}^\infty {p \choose k} x^k .
\]
Hence when $|\sigma|\leq t$,
\[
(t+\sigma)^p+(t-\sigma)^p = 2t^p\sum_{k\text{ even}} {p\choose k}\left(\frac{\sigma}{t}\right)^k.
\]
Thus
\[
\|A\|_{S_p}^p = 2\sum_{k\text{ even}} \binom{p}{k}t^{p-k}\|M\|_k^k.
\]
The existence of a gap in $\|A\|_{S_p}^p$ follows immediately from properties (i) and (ii) above.
\end{proof}
We remark that all lower bounds in Theorem~\ref{thm:PSD_lb} even hold for sparse matrices,  since the hard instances are sparse. The lower bounds for non-integers $p$ and even integers $p$ are strengthenings of the same lower bounds in~\cite{li2016approximating}, and are almost tight and tight up to polylogarithmic factors, respectively.

\section{Row-Order Model: Lower Bounds}
\label{sec:RowOrder}

First we discuss lower bounds for estimating Schatten norms in the row-order model. Suppose that $G$ is a graph with $n$ nodes and $m=O(n)$ edges. Let $M\in \R^{m\times n}$ be the incidence matrix of $G$ and $L\in\R^{n\times n}$ be the Laplacian matrix of $G$, then $L = M^T M$ and thus $\|M\|_{S_p}^p = \|L\|_{S_{p/2}}^{p/2}$. Similarly to the approach in \cite{li2016approximating}, we shall need a lower bound on distinguishing two families of graphs, while some matrix derived from the graph has different Schatten norms in the two cases. The lower bound on distinguishing graphs we shall use is due to Kogan and Krauthgamer~\cite{KK15} 
based on the Boolean Hypermatching Problem~\cite{VY11}, defined as follows.
\begin{proposition}[\cite{KK15}]
\label{prop:cycles}
Let $t\geq 2$ be an integer,
and let $G$ be an undirected $2$-regular graph on $n$ nodes consisting of either (a) vertex-disjoint $(2t+1)$-cycles or (b) vertex-disjoint $(4t+2)$-cycles. 
Every randomized one-pass insertion-only streaming algorithm that, with probability at least $2/3$, determines whether $G$ is of type (a) or type (b) must use $\Omega_t(n^{1-1/t})$ bits of space.
\end{proposition}

The next lemma shows that the Laplacian matrix has different Schatten $p$-norms between the two cases in the hard instance. 
\begin{lemma}\label{lem:schatten_gap}
Suppose that $t\geq 2$ is an integer and $p > 0$ is not an integer. 
Let $G$ be a graph as in Proposition~\ref{prop:cycles}, then the Schatten $p$-norm of the Laplacian matrix of $G$ is different by a constant factor $c(t,p) \neq 1$ between the two types.
\end{lemma}
\input{cycle_gap_proof}

The next theorem follows easily by combining
Proposition~\ref{prop:cycles} and Lemma~\ref{lem:schatten_gap}.

\begin{theorem}\label{thm:non-int-lb}
Suppose that $t\geq 2$ is an integer and $p > 0$ is not an even integer. 
Every randomized streaming algorithm that with probability at least $2/3$
estimates $\|A\|_{S_p}^p$ within factor $1+\epsilon$,  
for $\epsilon<\epsilon_0(t,p)$,  
when the input $A\in \R^{n\times n}$ is sparse and given in row-order model, 
must use $\Omega_t(n^{1-1/t})$ bits of space.
\end{theorem}
\begin{proof}
Let $A$ be the incidence matrix of $G$ in Proposition~\ref{prop:cycles}. Since $G$ has exactly $n$ edges, the size of $A$ is exactly $n\times n$. In the streaming model for $G$, each update describes an edge, which corresponds to a row of $A$. Thus a stream of $G$ corresponds to a stream of $A$ in row-order model. By Lemma~\ref{lem:schatten_gap}, a Schatten-norm algorithm can distinguish the type of $G$, and the lower bound therefore follows from Proposition~\ref{prop:cycles}.
\end{proof}

The theorem above gives a nearly tight bound for estimating the Schatten $p$-norm for sparse matrices and $p\notin 2\Z$. For $p\in 2\Z$ we have the following theorem.

\begin{theorem}\label{thm:even-int-lb}
Suppose that $t\geq 2$ is an integer and $p \geq 4$ is an even integer. Every randomized streaming algorithm that estimates the Schatten $p$-norm of the input matrix up to a constant factor (depending on $p$) with probability $\geq 2/3$ in the row-order model must use $\Omega(n^{1-4/p})$ bits of space. This lower bound holds even for multi-pass algorithms.
\end{theorem}

\begin{proof}
We reduce the problem to the communication complexity of multiparty \textsc{Set-Disjointness} \cite{gro09}. Suppose there are $k=2n^{2/p}$ players. Each player is given a set in $\{1,\dots,n\}$. Let $A$ be an empty matrix and we shall show how to construct $A$ according to the input of the \textsc{Set-Disjointness} problem. For each element $j$ in each player's set, we add a row $e_j$ to $A$, where $e_j$ is the $j$-th row of the $n\times n$ identity matrix. With high probability the hard instance of multiparty \textsc{Set-Disjointness} has $m\leq n$ elements, and thus $A$ will have $m$ rows. By padding we may assume that $A$ is $n\times n$, and is clearly given in the row-order model.

When the players' sets are disjoint, it is clear that all singular values of $A$ are $1$ and thus $\|A\| = m \leq n$. When the players' set have a common element, there is a singular value of $\sqrt k$ and hence $\|A\|_{S_p}^p \geq k^{p/2} = 2^{p/2}n$. Therefore $\|A\|_{S_p}^p$ is different by a constant factor in the two cases.

The communication complexity lower bound of unrestricted protocols for \textsc{Set-Disjointness} is $\Omega(n/k)$ bits, which implies that the streaming lower bound for estimating Schatten $p$-norm is $\Omega(n/k^2) = \Omega(n^{1-4/p})$ bits, even for multi-pass algorithms.
\end{proof}

As discussed in Introduction, Theorem~\ref{thm:non-int-lb} is asymptotically tight up to logarithmic factors for $p\in 4\Z$. For the remaining case $p\equiv 2\pmod{4}$, we  present an algorithm using $\tilde O(n^{1-4/(p+2)})$ space in Section~\ref{sec:row_order_ub}, leaving a slight polynomial gap from the lower bound of $\Omega(n^{1-4/p})$.


%% file: cycle_gap_proof.tex
\begin{proof}
Let $m=4t+2$. To prove the lemma it suffices to show a gap in the Schatten-$p$ norm of the Laplacian matrix between two $(m/2)$-cycles and one $m$-cycles. Let $L_m$ denote the Laplacian matrix of an $m$-cycle. Since $L_m$ is circulant, its eigenvalues (and thus singular values since $L$ is PSD) are given by the following explicit expression:
\[
\sigma_{m,j} = 2 - \omega_m^j - \omega_m^{j(m-1)},\quad j = 0,\dots,m-1
\]
where
\[
\omega_m = e^{2\pi i\frac{\pi}{m}}.
\]
Thus
\[
\|L_m\|_{S_p}^p = 
\begin{cases}
2\sum_{i=1}^{\lfloor m/2\rfloor} \sigma_{m,j}^p, & m\text{ is odd};\\
4^p + 2\sum_{i=1}^{m/2-1} \sigma_{m,j}^p, & m\text{ is even}.
\end{cases}
\]
When $m$ is an even integer, the eigenvalues of $L_{m/2}$ are eigenvalues of $L_{m}$, more specifically, $\sigma_{m/2,j} = \sigma_{m,2j}$. It follows that
\[
2\|L_{m/2}\|_{S_p}^p - \|L_{m}\|_{S_p}^p = 2\sum_{j=1}^{\frac n2-1} (-1)^j\sigma_{m,j}^p - 4^p,
\]
where we used the fact that $m=4t+2$ in our setting and thus $m/2=2t+1$ is odd. Note that
\[
\sigma_{m,j} = 2 - 2\cos\frac{2j\pi}{m} = 4\sin^2\frac{j\pi}{m},
\]
we have that
\[
2\|L_{m/2}\|_{S_p}^p - \|L_{m}\|_{S_p}^p = 2 \cdot 4^p\sum_{j=1}^{\frac n2-1} (-1)^j\sin^{2p}\frac{j\pi}{m} - 4^p.
\]
Our goal is therefore to show that
\[
\sum_{j=1}^{\frac n2-1} (-1)^j\sin^{2p}\frac{j\pi}{m} \neq \frac{1}{2}.
\]

Consider the Fourier cosine series expansion
\begin{align*}
\sin^{2p}\frac{j\pi}{m} = \frac{1}{2^{2p}}\frac{\Gamma(2p+1)}{\Gamma(p+1)^2}  
+ \frac{1}{2^{2p-1}}\sum_{k=1}^{\infty}\frac{ (-1)^k\Gamma(2p+1)}{\Gamma(p+k+1)(p-k+1)}\cos\left(2kj\frac{\pi}{m}\right),
\end{align*}
where the Fourier coefficient can be obtained by using Binomial Theorem and Gauss Theorem for hypergeometric functions ${}_2F_1$ to evaluate the following integral (cf.\ Exercise 44 on p123 of~\cite{AAR99})
\[
\int_{-\pi/2}^{\pi/2} (1-e^{2i\theta})^{p-k}(1-e^{-2i\theta})^{p+k}d\theta.
\]
Next, observe that
\begin{equation}\label{eqn:alternating_sum_2}
\sum_{j=1}^{\frac{m}{2}-1} (-1)^j \cos\left(2kj\frac{\pi}{m}\right) = \begin{cases}
0, & k\text{ is even};\\
\frac{m}{2}-1, & k\equiv \frac{m}{2}\!\!\!\!\pmod m;\\
-1, & \text{otherwise}.
\end{cases}
\end{equation}
The problem reduces to evaluate
\[
S := \frac{1}{2^{2p-1}} \bigg\{\sum_{\substack{\text{odd }k\\ k\not\equiv \frac{m}{2}-1(\!\bmod m)}} \!\!\!\!\!\!\!\! \gamma_p(k) - \left(\frac{m}{2}-1\right)\!\!\!\!\sum_{\substack{k\equiv \frac{m}{2}-1(\!\bmod m)}} \!\!\!\!\!\!\!\!\gamma_p(k) \bigg\},
\]
where
\[
\gamma_p(k) = \frac{\Gamma(2p+1)}{\Gamma(p+k+1)\Gamma(p-k+1)}.
\]
Observe that $\gamma_p(k) > 0$ for $k \leq \lceil p\rceil$, $\gamma_p(\lceil p\rceil+1) < 0$ and $\gamma_p(k)$ has alternating signs for $k\geq \lceil p\rceil+1$. 

When $\lceil p\rceil$ is even, it holds that $\gamma_p(k) < 0$ for $k \equiv m/2-1(\bmod\ m)$ and thus
\[
S > \frac{1}{2^{2p-1}} \sum_{\text{odd }k} \gamma_p(k);
\]
when $\lceil p\rceil$ is odd, it holds that $\gamma_p(k) > 0$ for $k \equiv m/2-1(\bmod\ m)$ and thus
\[
S < \frac{1}{2^{2p-1}} \sum_{\text{odd }k} \gamma_p(k).
\]
The result follows immediately once the following identity is established:
\begin{align}\label{eqn:final_identity}
\frac{1}{2^{2p-1}} \sum_{\text{odd }k} \gamma_p(k) = \frac{1}{2^{2p-1}} \sum_{\text{odd }k} \frac{\Gamma(2p+1)}{\Gamma(p+k+1)\Gamma(p-k+1)} = \frac{1}{2}.
\end{align}
Consider the integral representation
\begin{align*}
\frac{\Gamma(2p+1)}{\Gamma(p+k+1)\Gamma(p-k+1)} 
= \frac{1}{2\pi i}\int_{-\infty}^{(0+)} t^{-(p-k)-1}(1-t)^{-(p+k)-1} dt,
\end{align*}
where the contour integral goes from the upper edge of the negative real axis from $-\infty$ to $0$, then goes clockwise around $0$, and returns to $-\infty$ along the lower edge of the negative real axis. Summing under the integral yields that
\begin{align*}
&\quad\ \sum_{\text{odd }k} \gamma_p(k) \\
&= \frac{1}{2\pi i}\int_{-\infty}^{(0+)} \frac{1}{(1-2t)t^p(1-t)^p} dt \\
&= \frac{\sin(p\pi)}{\pi} {}_2F_1\left(\begin{array}{c}1,1-p\\1+p\end{array};-1\right)\frac{\Gamma(2p)\Gamma(1-p)}{\Gamma(1+p)} \\
&= \frac{\sin(p\pi)}{\pi} \cdot \frac{\Gamma(1+p)\Gamma(\frac{3}{2})}{\Gamma(2)\Gamma(\frac{1}{2}+p)} \cdot \frac{\Gamma(2p)\Gamma(1-p)}{\Gamma(1+p)}\\
&= \frac{\sin(p\pi)}{\pi} \cdot \frac{\sqrt\pi/2}{\Gamma(\frac12\!+\!p)}\cdot \frac{2^{2p-1}}{\sqrt\pi}\Gamma(p)\Gamma\!\left(p+\!\frac12\right)\Gamma(1\!-\!p)\\
&= 2^{2p-2},
\end{align*}
where we used the integral representation of hypergeometric function ${}_2F_1$ (Equation (2.3.17) in \cite{AAR99}) for the second equality, Kummer's identity (\cite[Corollary 3.1.2] {AAR99}) for the third equality, Legendre's duplication formula (\cite[Theorem 1.5.1]{AAR99}) for the fourth and Euler Reflection Formula (\cite[Theorem 1.2.1]{AAR99}) for the last equality. This establishes \eqref{eqn:final_identity}.
\end{proof}

%% file: row_order_ub.tex
\section{Row-Order Model: Algorithm For Even \mathp} 
\label{sec:row_order_ub}
\input{row_order_algorithm}
In this section, we present an algorithm which estimates the Schatten $p$-norm (where $p\equiv 2\pmod{4}$ is an integer) of $n\times n$ sparse matrices in row-order model using $\tilde O(n^{1-4/(p+2)})$ bits of space. The following algorithm is in a similar flavour to the algorithm in \cite{li2016approximating}, where the Precision Sampling structure was used to sample rows of a matrix proportionally to their row norms, and we shall omit such details in this section. Since we are reading $A$ in row-order model, we can sample and obtain rows of $A$ exactly with weighted reservoir sampling~\cite{ES06}, but we shall use Precision Sampling to sample rows of $A^TA$.

\begin{theorem}\label{thm:row_order_ub}
Suppose that $p=4k+2$ for some integer $k\geq 1$ and $A\in \R^{n\times n}$ is a sparse matrix given in one-pass row-order model. Algorithm~\ref{alg:row_order} returns $Y$ such that $(1-\epsilon)\|A\|_{S_p}^p \leq Y\leq (1+\epsilon)\|A\|_{S_p}^p$ with probability $\geq 2/3$, using space $O_p(n^{1-\frac{4}{p+2}}\poly(1/\epsilon,\log n))$.
\end{theorem}

\begin{proof}
The analysis is similar to \cite{li2016approximating}. Let $B=A^TA$, $L=\|B\|_F^2$ and $Z=\|A\|_F^2$. For a matrix $M$ we shall denote its $i$-row by $M_i$. We also denote by $\tilde B_i$ the approximation recovered by the algorithm to $B_i$. For notational convenience, we also define $K_1=\cdots=K_k=K$ and $K_{k+1}=V$.

Next, we define for $s=1,\dots,k$
\begin{align*}
\tau_s(i) &= \begin{cases}
						1, & i \in K;\\
						L/\|B_i\|_2^2, & i\in I_s\setminus K,
					\end{cases}
\\
\tilde\tau_s(i) &= \begin{cases}
									1, & i \in K;\\
									L'/\|\tilde B_i\|_2^2, & i\in I_s\setminus K
							 \end{cases}
\end{align*}
and
\[
\tilde\tau_{k+1}(i) = \tau_{k+1}(i) = \begin{cases}
						1, & i \in V;\\
						Z/\|A_i\|_2^2, & i\in I_{k+1}\setminus V.
					\end{cases}
\]
We further define
\begin{align*}
X(i_1,\dots,i_t)=
\prod_{j=1}^{k} \langle B_{i_j},B_{i_{j+1}}\rangle\langle B_{i_{j+1}},A_{i_{k+1}}\rangle\langle A_{i_{k+1}},B_{i_1}\rangle\cdot\tau_1(i_1)
\cdots\tau_{k+1}(i_{k+1}),
\end{align*}
and
\begin{align*}
\tilde X(i_1,\dots,i_t)=
\prod_{j=1}^{k} \langle \tilde B_{i_j},\tilde B_{i_{j+1}}\rangle\langle \tilde B_{i_{j+1}}, A_{i_{k+1}}\rangle\langle  A_{i_{k+1}},\tilde B_{i_1}\rangle \cdot\tilde\tau_1(i_1)
\cdots\tilde\tau_{k+1}(i_{k+1}).
\end{align*}
Since $B=A^TA$ is PSD, it holds that
\begin{align*}
\|A&\|_{S_p}^p = \tr(B^{p/2}) \\
				&= \sum_{i_1} e_{i_1}^T B \cdot \underbrace{(B^TB)\cdots(B^T B)}_{k-1\text{ times}} \cdot B \cdot Be_{i_1} \\
				&= \sum_{i_1} B_{i_1} \cdot \underbrace{(B^TB)\cdots(B^T B)}_{k-1\text{ times}} \cdot A^TA \cdot B_{i_1}^T \\
				&= \sum_{i_1,\dots,i_{k+1}} B_{i_1} (B^T_{i_2} B_{i_2})\cdot (B^T_{i_k} B_{i_k}) (A^T_{i_{k+1}} A_{i_{k+1}}) B_{i_1}^T \\
				&= \sum_{i_1,\dots,i_{k+1}} \langle B_{i_j},B_{i_{j+1}}\rangle\langle B_{i_{j+1}},A_{i_{k+1}}\rangle\langle A_{i_{k+1}},B_{i_1}\rangle.
\end{align*}
Our estimator is
\begin{equation}\label{eqn:estimator}
Y = \sum_{i\in I_1,\dots,i_{k+1}\in I_{k+1}} \frac{1}{T^{\sigma(i_1,\dots,i_{k+1})}}\tilde X(i_1,\dots,i_{k+1}),
\end{equation}
where
\[
T^{\sigma(i_1,\dots,i_{k+1})} = \left| \left\{1\leq s\leq k+1: i_s\not\in K_s \right\} \right|
\]
Following a similar analysis to that in~\cite{li2016approximating}, we have that
\[
\left|\E Y - \|A\|_{S_p}^p\right|\leq \epsilon\|A\|_{S_p}^p,
\]
where we have crucially used the fact that $A$ and $A^T A$ are sparse matrices.
The variance bound is similar, too. The covariance terms are sums over $i_1,\dots,i_{k+1},i_1',\dots,i_{k+1}'$ and we split them into two kinds depending on whether $i_{k+1}=i_{k+1}'$. Eventually we shall have
\begin{align*}
\E Y^2 - (\E Y)^2 &\lesssim \sum_{r=1}^k \frac{1}{T^r} \|B\|_F^{2r} \|A\|_{S_{2p-4r+4}}^{2p-4r+4}  + \sum_{r=1}^{k+1} \frac{1}{T^r} \|B\|_F^{2(r-1)}\|A\|_F^2 \|A\|_{S_{2p-4r+2}}^{2p-4r+2} 
&\lesssim \sum_{r=1}^{k+1} \frac{1}{T^r}n^{r-\frac{4r-2}{p}} \|A\|_{S_p}^{2p},
\end{align*}
which implies that
\[
\E Y^2 - (\E Y)^2\leq \epsilon^2 \|A\|_{p}^{2p}
\]
if the constant $C$ in $T=Cn^{1-1/(k+1)}/\epsilon^2$ is large enough.
\end{proof}


%% file: row_order_algorithm.tex
\begin{algorithm*}[th]
\caption{Algorithm for $p=4k+2$ and sparse matrices in row order model}\label{alg:row_order}
\begin{algorithmic}[1]
\Statex Assume that matrix $A$ has at most $O(1)$ non-zero entries per row and per column and is given in row order model and that $p=4k+2$ for some integer $k\geq 1$.
\State $T\gets \Theta(\epsilon^{-2}n^{1-1/(k+1)})$
\State Maintain a sketch for estimating $\|A^TA\|_F^2$ and obtain an $(1+\epsilon)$-approximation $L'$
\State $Z\gets \|A\|_F^2$ \Comment{\parbox[t]{.35\linewidth}{Can be computed exactly in row-order model\vspace{2mm}}}
\State $K\gets$ set of indices of rows of $A^T\!A$ with norm $\geq\!\! \sqrt{L'/(10T)}$ \Comment{\parbox[t]{.35\linewidth}{\textsc{Count-Sketch}, see~\cite{li2016approximating}\vspace{2mm}}}
\State $V\gets$ set of indices of rows of $A$ with norm $\geq \sqrt{Z/(10T)}$ \Comment{\parbox[t]{.35\linewidth}{Maintaining $10T$ rows of largest norm}}
\For{$s=1,\dots,k$}
\State Sample $T$ rows of $A^T A$ proportionally to row norm
\Comment{\parbox[t]{.35\linewidth}{
Precision sampling, see~\cite{li2016approximating}}}
\State Obtain approximation to the sampled rows
\Comment{\parbox[t]{.35\linewidth}{
By-product of precision sampling, see~\cite{li2016approximating}}}
\State $I_s\gets$ the set of the indices of the sampled rows
\State $I_s\gets I_s\cup K$
\EndFor
\State Sample $T$ rows of $A$ proportionally to row norm \Comment{\parbox[t]{.35\linewidth}{Reservoir sampling}}
\State $I_{k+1}\gets $ the set of the indices of the sampled rows
\State $I_{k+1}\gets I_{k+1}\cup V$
\State Return $Y$ as defined in \eqref{eqn:estimator}
\end{algorithmic}
\end{algorithm*}

%% file: appendix.tex
\input{appendix_schatten}

%% file: appendix_schatten.tex


\section{Proof of  Proposition~\ref{prop:gaussian sketch for sp integer}}
\label{sec:proof gaussian sketch for sp integer}
\begin{proof}
Using the identity $\tr(MM^T) = \tr(M^TM)$ we have
\begin{align*}
X = \tr\left(G_1 AG_2^TG_2 A\cdots G_p^TG_pA \cdot G_1^T\right) 
 = \tr\left(G_1^T\cdot G_1 AG_2^TG_2A\cdots G_p^TG_p A\right).
\end{align*}
By linearity of trace, expectation and matrix product, 
and by the fact that 
$
 \EX[G_i^TG_i] = I_{n\times n}
$
for all $i\in[p]$, 
we have 
\begin{align*}
  \EX X
  &= \EX \tr\left(G_1^T\cdot G_1 AG_2^TG_2A\cdots G_p^TG_p A \right)  \\
  &= \EX \tr\left( I_{n\times n} AG_2^TG_2A\cdots G_p^TG_p A \right)  \\
  &= \cdots = \tr(A^p).
\end{align*}

It remains to bound the variance of $X$.  
Without loss of generality we can assume that $A$ is 
a diagonal matrix $\diag(\lambda_1, \lambda_2, \dots, \lambda_n)$, 
where $\lambda_1\ge \cdots \ge\lambda_n$. 
Indeed, in the case of a general symmetric $A$, we can write 
$A=U\Lambda U^T$ for an orthonormal matrix $U$ and a diagonal matrix $\Lambda$.
Then $G_i A G_{i+1}^T = (G_i U)\Lambda (G_{i+1}U)^T$,
and the matrices $\set{G_iU}_{i\in[p]}$ have the same joint 
distribution as $\set{G_i}_{i\in[p]}$,
hence $\Var(X)$ would not change if $A$ is replaced with $\Lambda$.

Let us write $G_i=(g^{(i)}_1, g^{(i)}_2, \dots, g^{(i)}_n)$, 
where each $g^{(i)}_j\in\bbR^{t}$ is a column vector.
It is easily verified that 
\[
  X
  = \!\!\! \sum_{i_1, i_2, \dots, i_p\in[n]} \!\!\!\!\!\!
    \lambda_{i_1}\lambda_{i_2}\cdots \lambda_{i_p}
    \langle g^{(1)}_{i_1}, g^{(1)}_{i_2}\rangle
    \langle g^{(2)}_{i_2}, g^{(2)}_{i_3}\rangle
    \cdots \langle g^{(p)}_{i_p}, g^{(p)}_{i_1}\rangle.
\]
Indeed, first write 
\[(G_1^TG_1 A)_{i_1,i_2}
 = \sum_{k\in[t]} (G_1^T)_{i_1,k} (G_1)_{k,i_2} A_{i_2,i_2}
 = \tuple{g^{(1)}_{i_1}, g^{(1)}_{i_2}} \lambda_{i_2},\]
and then expand the trace in\\
$X = \tr\left(G_1^TG_1 A\cdot G_2^TG_2A \cdots G_p^TG_p A\right)$
using all closed walks $(i_1,i_2,\dots, i_p)\in[n]^p$.

It is not difficult to verify that for all $j\neq j'\in[n]$ 
and $i_1,i_2,i'_1,i'_2\in[p]$,
\begin{align}
  &\EX[ \langle g_{i_1}^{(j)}, g_{i_2}^{(j)}\rangle ] 
  = \indice{i_1=i_2}, 
  \label{eqn:sqr_exp0}
  \\
  &\EX[ \tuple{ g_{i_1}^{(j)}, g_{i_2}^{(j)} } \tuple{ g_{i'_1}^{(j')}, g_{i'_2}^{(j')} } ]
  = \indice{i_1=i_2, i'_1=i'_2}.
  \label{eqn:sqr_exp1}
  \\
  &\EX[ \tuple{ g_{i_1}^{(j)}, g_{i_2}^{(j)} } \tuple{ g_{i'_1}^{(j)}, g_{i'_2}^{(j)} } ]
  = \indice{(i_1,i_1')=(i_2,i_2')} + \tfrac{1}{t} \indice{(i_1,i_2) =(i_1',i_2')} 
    + \tfrac{1}{t} \indice{(i_1,i_2) =(i_2',i_1')} . 
    \label{eqn:sqr_exp2}
\end{align}
Notice that in the last equation, 
the events in the three indicators are not disjoint,
and when $i_1=i'_1=i_2=i'_2$ the righthand-side evaluates to $1+2/t$.

We proceed to bound $\Var(X) \leq \EX[X^2]$.
Denoting $I=(i_1, i_2, \dots, i_p)\in[n]^p$ with the convention $i_{p+1}:=i_1$,
and similarly for $I'$, we can write
\begin{align} \label{eq:Xsqr} 
  X^2 
  = \left( 
    \sum_I \prod_{j\in[p]} \lambda_{i_j} \tuple{ g^{(j)}_{i_j}, g^{(j)}_{i_{j+1}} }
    \right)^2
  = \sum_{I, I'} \prod_{j\in[p]}\lambda_{i_j}\lambda_{i_j'}
    \tuple{ g_{i_j}^{(j)}, g_{i_{j+1}}^{(j)} } \tuple{ g_{i_j'}^{(j)}, g_{i_{j+1}'}^{(j)} }.
\end{align}

\begin{figure*}
\input{gaussian_variance}
\caption{An example of a non-zero variance term ($p=9$): $i_1=\cdots = i_5 = i_1'=i_2'\cdots = i_5'$,
$i_6=i_6'$, $i_7 = i_7'$, $i_8=i_9=i_8'=i_9'$ and $i_1, i_6, i_7, i_8$ are distinct. The term of the eigenvalues in the variance expression is $\lambda_{i_1}^{10}\lambda_{i_6}^2\lambda_{i_7}^2\lambda_{i_8}^4$.
\label{fig:variace term}}
\hrule
\end{figure*}
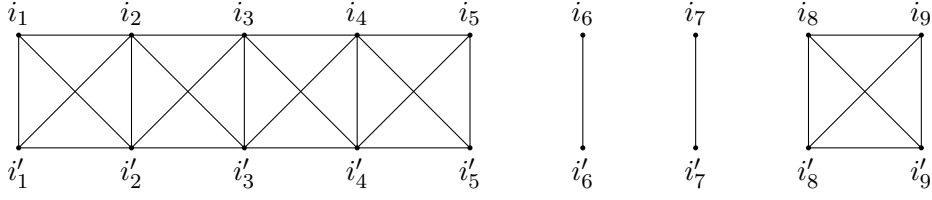

We can represent each term of $X^2$ (a fixed choice for $I,I'$) by a diagram 
(see an example in Figure~\ref{fig:variace term}). 
Each node in the diagram represents an index $i_j$,
and each square corresponds to a factor of the form
$
\tuple{ g_{i_j}^{(j)}, g_{i_{j+1}}^{(j)} } \tuple{ g_{i_j'}^{(j)}, g_{i_{j+1}'}^{(j)} }
$.
A line connecting two nodes represents that the respective indices are equal.
Notice that for each square, if a vertical line exists, 
then both vertical lines must exist, 
otherwise the expectation of this square is zero, 
and it has no contribution to $\EX[X^2]$.
Thus, for a non-zero diagram, if it has at least one vertical line, 
then it actually has all possible vertical lines. 
Diagrams with no vertical lines can be non-zero diagrams only if they
are made entirely by horizontal cross-lines and parallel lines,
which we call the \emph{trivial diagrams}, 
and they correspond to the coefficient of $\lambda_i^p\lambda_j^p$ for $i\neq j$. 
Each non-trivial diagram corresponds to an \emph{integer partition} of $p$
(i.e., a way of writing the integer $p$ as the sum of positive integers,
with the order of the summands/parts having no significance),
but we should account also for permutations and cyclic shifts on the parts. 
Given an integer partition $[p_1, p_2, \dots, p_z]$ of $p$, 
we write it as $(p_1^{(z_1)}, p_2^{(z_2)}\cdots p_{t'}^{(z_{t'})})$, 
where $p_1\ge p_2\ge \cdots \ge p_{t'}$ are the distinct parts (or part sizes),
and $z_i$ counts how many parts are equal to $p_i$. 
Then the number of different diagrams 
for a given integer partition $[p_1, p_2, \dots, p_z]$ is 
\[
  C_{[p_1, p_2, \dots, p_z]}=\frac{t'!p}{z_1!z_2!\cdots z_{t'}!}.
\] 
Observe that this number is upper bounded by a constant $M_p$ determined only by $p$.
Each integer partition of $p$ corresponds to a monomial of the eigenvalues. 
A connected component in the diagram corresponding to a power of the eigenvalue, and this power is just the size of that component. 
For each connected component, the total number of indices is an even number
because of the vertical lines.
For a single square, the coefficient is given by 
Equations \eqref{eqn:sqr_exp1}-\eqref{eqn:sqr_exp2}. 
Using the diagram representation, we can calculate
\begin{equation}\label{eq:Esquares}
\begin{aligned} 
\EX\left(\ \compsq\ \right) = 1+\frac{2}{t};
\qquad
\EX\left(\ \updownsq\ \right) = 1;
\\
\EX\left(\ \leftrightsq\ \right) 
= \EX\left(\ \crosssq\ \right) 
= \frac{1}{t}.
\end{aligned}
\end{equation}
All other diagrams either do not exist in the expansion of $X^2$, 
or have a zero expectation.
Diagrams corresponding to the same partition of $p$ have the same coefficient. Since for each complete square there is a factor $1+2/t$, 
and for each incomplete square there is a factor $1/t$, 
the coefficient for a partition $[p_1, p_2, \dots, p_z]$ with $z>1$ parts is 
\[
  Z_{[p_1, p_2, \dots, p_z]}
  = \frac{1}{t^{z}}\left(1+\frac{2}{t}\right)^{p-z}.
\]
For non-trivial diagrams (i.e., have vertical lines) 
with $z>1$ (i.e., excluding the completely connected graph) 
we collect all such terms as $X_1$ and bound their expectation by
\begin{align} \label{eq:X_1}
  \EX[ X_1]
  \le \sum_{[p_1, p_2, \dots, p_z]} \frac{M_p}{t^z}
      \sum_{i_1, i_2, \dots,i_z\in[n]} \lambda_{i_1}^{2p_1} \lambda_{i_2}^{2p_2} 
         \cdots\lambda_{i_z}^{2p_z} .
\end{align}
For non-trivial diagrams and $z=1$, 
there cannot be any incomplete square,
and we can compute the expression explicitly,
\begin{align*}
  \EX \bigg[\sum_{i\in[n]} \lambda_{i}^{2p}\prod_{j\in [p]}\langle g_{i}^{(j)}, g_{i}^{(j)}\rangle\langle g_{i}^{(j)}, g_{i}^{(j)}\rangle\bigg]
  &= \left(1+\frac{2}{t}\right)^p\sum_{i=1}^n\lambda_i^{2p}\\
  &= 3^p \norm{A}_{S_{2p}}^{2p} .
\end{align*}
For trivial diagrams (no vertical lines), 
we collect the terms as $X_2$ and bound their expectation by
\begin{equation} \label{eq:X_2}
\begin{aligned}
  &\EX[X_2]\nonumber\\ 
  &~\le \sum_{i\neq k\in[n]} \lambda_{i}^{p}\lambda_k^p
\sum_{z=0}^p  \binom{p}{z} \EX\bigg(\ \crosssq\ \bigg)^z
\EX\bigg(\ \updownsq\ \bigg)^{p-z} \\
  &~\le 2^p 
      \sum_{i\neq k\in[n]} \lambda_{i}^{p}\lambda_k^p\\
  &~\le M_p' \norm{A}_{S_p}^{2p},
\end{aligned}
\end{equation}
where $M_p'$ is a constant that depends only on $p$. 

We now turn to bounding $\EX[X_1]$ using \eqref{eq:X_1}.
For each integer partition $[p_1, p_2, \dots, p_z]$ of $p$ with $z>1$ parts,
\begin{align*}
 \sum_{i_1, i_2, \dots, i_z\in[n]}
   \lambda_{i_1}^{2p_1}\lambda_{i_2}^{2p_2}\cdots\lambda_{i_z}^{2p_z}
 &= \bigg(\sum_{i_1\in[n]} \lambda_{i_1}^{2p_1} \bigg)
   \cdots
   \bigg(\sum_{i_z\in[n]} \lambda_{i_z}^{2p_z} \bigg)\\
 &= \prod_{j=1}^z \norm{A}_{S_{2p_j}}^{2p_j}. 
\end{align*}
Let $z'$ be the number of parts with $2p_j\le p$.
Clearly, $z'\ge z-1$, since at most one part can have $p_j\ge p/2$. 
Consider first the case $z'=z$.
It is well-known (via an application of \Holder's inequality) 
that $\norm{x}_{q} \leq \norm{x}_{r} \leq n^{1/r-1/q} \norm{x}_{q}$
holds for all $x\in\R^n$ and $1\le r\le q$.
This comparison of norms applies also to the Schatten norms of $A$
(viewed as $n$-dimensional norms of the eigenvalues of $A$), 
proves that $\norm{A}_{S_{2p_j}}^{2p_j} \leq n^{1/(2p_j) - 1/p} \norm{A}_{S_p}$.
We thus obtain
\begin{align} \label{eq:z}
  \prod_{j=1}^z \norm{A}_{S_{2p_j}}^{2p_j} 
  \le \prod_{j=1}^z \left( n^{1/(2p_j) - 1/p} \norm{A}_{S_p} \right)^{2p_j}
  =   n^{z-2} \norm{A}_{S_p}^{2p}.
\end{align}
In the case $z'=z-1$, there is a unique $j^*$ such that $p_{j^*}>p/2$,
and therefore $z\le (p-p_{j^*})+1\le \lfloor p/2\rfloor + 1$.
For $j\neq j^*$ we can use the comparison of Schatten norms as above,
and for $j=j^*$ we simply use $\norm{A}_{S_{2p_j}} \leq \norm{A}_{S_{2p}}$.
We thus obtain
\begin{equation}\label{eq:zminus1}
\begin{aligned} 
  \prod_{j=1}^z \norm{A}_{S_{2p_j}}^{2p_j} 
  &\le \norm{A}_{S_{2p}}^{2p_{j^*}} 
       \prod_{j\neq j^*} \left(n^{1/(2p_j) - 1/p} \norm{A}_{S_p} \right)^{2p_{j}} \\
  &\le n^{z-1 - 2(p-p_{j^*}) / p} \|A\|_{S_p}^{2p} \\
  &\le n^{z-2z/p} \|A\|_{S_p}^{2p} .
\end{aligned}
\end{equation}
where the last inequality follows by 
$ 1+2(p-p_{j^*})/p
  \ge 1 + 2(z-1)/{p}
  = 2z/p + (1-2/p)
$.
%
With also the $z=1$ term considered, we have 
\[
  \Var(X)
  \le M_p''\!\!\left(1\!\!+\!\!\!\!\sum_{z=2}^{\lfloor p/2\rfloor + 1}\left(\frac{n^{1-2/p}}{t}\right)^{\!z}\!\!+\!\!\sum_{z=2}^{p}\left(\frac{n^{1-2/z}}{t}\right)^{\!z}\right)\!\!\|A\|_{S_p}^{2p}.
\]
where $M_p''$ is a constant depends only on $p$.
This completes the proof of 
Proposition~\ref{prop:gaussian sketch for sp integer}.
\end{proof}

\section{Proof of Lemma~\ref{cor:reduced size sketch for s norm}}
\label{sec:proof reduced size sketch for s norm}
\begin{proof}
We first argue that it suffices to prove the corollary 
under the assumption that the entries of $G_l$ are fully independent.
Indeed, each of the terms we need to calculate is the expectation of 
a polynomial of total degree at most $4$ in the random variables $G_{ij}$.
For example, the factor contains $G_l$ in a typical term  of $X^2$
is 
$
\langle g_{i_l}^{(l)} g_{j_l}^{(l)}\rangle
\langle g_{i_l'}^{(l)} g_{j_l'}^{(l)}\rangle
$
The expectation of such a polynomial when $G_l$'s entries are $4$-wise independent is exactly the same as when these entries are fully independent.

Assume henceforth that the entries of $G_l$ are mutually independent.  We  repeat the proof of Proposition~\ref{prop:gaussian sketch for sp integer},
except that when considering the square containing $(i_1, i_2)$, 
we replace $t$ with $t'$ in \eqref{eqn:sqr_exp2} and \eqref{eq:Esquares}.
In diagrams where this square is complete,
the contribution to $\EX[X^2]$, as given by \eqref{eq:Xsqr}, does not change.
When this square is incomplete, 
we replace $t$ by $t'$ in subsequent calculations 
like \eqref{eq:X_1} and \eqref{eq:X_2}.
The proof is otherwise identical,
but we kept the more precise bound obtained in \eqref{eq:zminus1}.
\end{proof}

\section{A Simple Proof for Sparse Sketch of Matrices With Non-Negative Entries}
\label{sec:simple proof for matrices with all positive entries}
 \begin{lemma}
 \label{lemma:property of sparse}
 Let $G=(g_1,g_2,\dots,g_n)\sim \calD_{t,n}$, then the following conditions hold.
 \begin{enumerate}
 \item for each $i\in[n]$, $\EX \langle g_i, g_i\rangle = 1$, $\EX[\langle g_i, g_i\rangle^2] = 1$;
 \item for each $i,j\in[n], i\neq j$, $\EX \langle g_i, g_j\rangle = 0$, $\EX[\langle g_i, g_j\rangle^2] = 1/t$;
 \item  for each $i,j, i', j'\in[n], \{i,j\}\neq \{i',j'\}, i\neq j, i'\neq j'$, $\EX\langle g_i, g_j\rangle = \EX[\langle g_i, g_j\rangle\langle g_{i'},g_{j'}\rangle] = 0$;
 \end{enumerate}
\end{lemma}
\begin{proof}
Property 1 follows immediately.
For $2$, $\EX \langle g_i, g_j\rangle =0$ and
\begin{align}
\EX \langle g_i, g_j\rangle^2 = \EX\left(\sum_{l} g_{i,l} g_{j, l}\right)^2
&= \sum_{l, k}
\EX(g_{i,l} g_{j, l}g_{i,k} g_{j, k}) \nonumber \\
&= \sum_{l=1}^tE(d_{i}^2d_{j}^2z_{i,l}z_{i,k})\nonumber \\
&= \sum_{l=1}^t\frac{1}{t^2}\nonumber =\frac{1}{t}.
\end{align}
For 3, we only need to consider the case when $\{i,j\}\cap\{i',j'\}\neq\emptyset$. Without loss of generality, assume $i=i'$, thus,
\begin{align*}
\EX \langle g_i, g_j\rangle\langle g_i, g_{j'}\rangle
&=
\sum_{l}E(g_{i,l}g_{j,l} g_{i,l}, g_{j',l})  \\
&\quad\ +\sum_{l\neq k}E(g_{i,l}g_{j,l} g_{i,k}, g_{j',k}) = 0,
\end{align*}
where we use that $g_{i,l}g_{i,k}=0$ when $l\neq k$.
\end{proof}
 
 The following lemma is a simple case that the variance of a sparse sketch is smaller than the Gaussian sketch. We will show in the next section that
 the sparse sketch is superior to the Gaussian sketch for every symmetric matrix. 
\begin{lemma}
 \label{lemma:sketch for unbiased JL for non-negative matrix}
Let $G_1\sim\calD_{t',n}$ and let $G_2, \dots G_{p}$ be independent copies of $\calD_{t,n}$, where $p\ge 2$ is an integer and $c_1, c_2$ are two absolute constants. Let $A$ be a symmetric matrix with all entries non-negative and $1\le t'\le t$. Let $X=\tr\left(G_1AG_2^TG_2AG_3\cdots G_{p}AG_1^T\right)$. Let $X'$ 
be a random variable obtained by replacing $G_i$ of $X$
by a column normalized gassian matrix of the same size. Then,
\[
\EX(X^2) \le \EX(X'^2).
\]
\end{lemma}
\begin{proof}
Let $J=\{j_1, \dots j_p\}\in[n]^p$ and
 $I=\{i_1, \dots i_p\}\in[n]^p$.
Define 
\begin{align*}
X_{I,J}&:=a_{i_p,j_1} a_{i_1, j_2}\cdots a_{i_{p-1}, j_p}\langle g_{j_1}^{(1)}, g_{i_1}^{(1)}\rangle\langle g_{j_2}^{(2)}, g_{i_2}^{(2)}\rangle\\
&\qquad\cdots 
\langle g_{j_p}^{(p)}, g_{i_p}^{(p)}\rangle.
\end{align*}
We now expand $X$ in a different form,
\vspace{-0.05in}
\[
X=\sum_{I,J}X_{I,J}.
\]
\vspace{-0.1in}
Thus,
\vspace{-0.1in}
\[
X^2=\sum_{I,J, I',J'}X_{I,J}X_{I',J'}.
\]
Define $X'_{I,J}$ analogously by replacing $g_i$ with Gaussian vectors. Since each $a_{i,j}\ge 0$,
with Proposition \ref{prop:gaussian sketch for sp integer} and Lemma \ref{lemma:property of sparse},
we immediately have that $\EX(X^2)\le \EX(X'^2)$. 
\end{proof}

The preceding lemma leads to the following theorem.
\begin{theorem}
\label{thm:fast update algorithm all positive}
For every integer $p\ge 2$, there exists a randomized one-pass streaming algorithm $\calA$ using space \\$O(n^{2-4/p} / \epsilon^2)$, and a $\lceil p/2\rceil$-pass streaming algorithm $\calB$ using space $O(n^{1-1/(p-1)} /\epsilon^2)$, given as input PSD matrix $A\in\bbR^{n\times n}$ with all entries non-negative, then
the output of the algorithms $\calA(A)$ and $\calB(A)$ satisfy
\[
\Pr[\calA(A)\in (1\pm \epsilon)\|A\|_{S_p}^p] \ge 0.99;
\]
and 
\[
\Pr[\calB(A)\in (1\pm \epsilon)\|A\|_{S_p}^p] \ge 0.99,
\]
where the probability is over the randomness of the algorithms. Both algorithms require $O(1/\epsilon^2)$ time to process each update in a pass. After the updates, $\calA$ requires
time $O(n^{(1-2/p)\omega}/\epsilon^2)$ to compute its output and
$\calB$ requires time $O(n^{(1-2/p)}/\epsilon^2)$, where $\omega<3$ is the matrix multiplication constant. 
For general input matrices $A$ of size $n\times m$ for  $m\le n$, if $A$ has all entries non-negative,  the above claim holds for even integers $p\ge 2$.
\end{theorem}

%% file: gaussian_variance.tex
\begin{center}
\begin{tikzpicture}[scale=0.75]
\draw (-8, 1) -- (0, 1) -- (0, -1) -- (-8, -1) -- cycle;
\draw (6, 1) -- (6, -1) -- (8, -1) -- (8,1) -- cycle;
\draw (-6, 1) -- (-6, -1);
\draw (-4, 1) -- (-4, -1);
\draw (-2, 1) -- (-2, -1);
\draw (2, 1) -- (2, -1);
\draw (4, 1) -- (4, -1);
\draw (-8, 1) -- (-6, -1);
\draw (-6, 1) -- (-4, -1);
\draw (-4, 1) -- (-2, -1);
\draw (-2, 1) -- (0, -1);
\draw (-8, -1) -- (-6, 1);
\draw (-6, -1) -- (-4, 1);
\draw (-4, -1) -- (-2, 1);
\draw (-2, -1) -- (0, 1);
\draw (6, -1) -- (8, 1);
\draw (6, 1) -- (8, -1);
\filldraw[black] (-8, 1) circle (1pt) node[anchor=south] {$i_1$};
\filldraw[black] (-6, 1) circle (1pt) node[anchor=south] {$i_2$};
\filldraw[black] (-4, 1) circle (1pt) node[anchor=south] {$i_3$};
\filldraw[black] (-2, 1) circle (1pt) node[anchor=south] {$i_4$};
\filldraw[black] (0, 1) circle (1pt) node[anchor=south] {$i_5$};
\filldraw[black] (2, 1) circle (1pt) node[anchor=south] {$i_6$};
\filldraw[black] (4, 1) circle (1pt) node[anchor=south] {$i_7$};
\filldraw[black] (6, 1) circle (1pt) node[anchor=south] {$i_8$};
\filldraw[black] (8, 1) circle (1pt) node[anchor=south] {$i_9$};
\filldraw[black] (-8, -1) circle (1pt) node[anchor=north] {$i_1'$};
\filldraw[black] (-6, -1) circle (1pt) node[anchor=north] {$i_2'$};
\filldraw[black] (-4, -1) circle (1pt) node[anchor=north] {$i_3'$};
\filldraw[black] (-2, -1) circle (1pt) node[anchor=north] {$i_4'$};
\filldraw[black] (0, -1) circle (1pt) node[anchor=north] {$i_5'$};
\filldraw[black] (2, -1) circle (1pt) node[anchor=north] {$i_6'$};
\filldraw[black] (4, -1) circle (1pt) node[anchor=north] {$i_7'$};
\filldraw[black] (6, -1) circle (1pt) node[anchor=north] {$i_8'$};
\filldraw[black] (8, -1) circle (1pt) node[anchor=north] {$i_9'$};
\end{tikzpicture}
\end{center}